\documentclass[12pt]{article}
\pdfoutput=1

\newlength{\abstractwidth}
\usepackage{amsmath}
\usepackage{amsfonts}
\usepackage{amssymb}
\usepackage{latexsym}

\usepackage{epsf}
\usepackage{color}
\usepackage{graphicx}
\usepackage{tikz}
\usepackage{dsfont}

\usetikzlibrary{arrows,shapes,positioning}
\usetikzlibrary{decorations.markings}
\usepackage[rightcaption]{sidecap}
\tikzstyle arrowstyle=[scale=1]
\tikzstyle directed=[postaction={decorate,decoration={markings,
    mark=at position .65 with {\arrow[arrowstyle]{stealth}}}}]
\tikzstyle reverse directed=[postaction={decorate,decoration={markings,
    mark=at position .65 with {\arrowreversed[arrowstyle]{stealth};}}}]
\usetikzlibrary{positioning}

\usepackage{hyperref}
\definecolor{darkred}{rgb}{0.8,0.1,0.1}
\hypersetup{colorlinks=true, linkcolor=darkred, citecolor=blue, linktoc=page}

\makeatletter\def\l@subsubsection#1#2{}%
\makeatother

\renewcommand{\thefootnote}{\fnsymbol{footnote}}
\renewcommand{\thanks}[1]{\footnote{#1}}
\newcommand{\starttext}{
\setcounter{footnote}{0}
\renewcommand{\thefootnote}{\arabic{footnote}}}

\newcommand{\bea}{\begin{eqnarray}}
\newcommand{\eea}{\end{eqnarray}}
\newcommand{\be}{\begin{eqnarray}}
\newcommand{\ee}{\end{eqnarray}}
\newcommand{\bal}{\begin{align}}
\newcommand{\eal}{\end{align}}
\newcommand{\bma}{\begin{matrix}}
\newcommand{\ema}{\end{matrix}}

\def\cA{{\cal A}}
\def\cB{{\cal B}}
\def\cC{{\cal C}}

\def\cG{{\cal G}}

\def\cI{{\cal I}}
\def\cJ{{\cal J}}
\def\cK{{\cal K}}
\def\cL{{\cal L}}
\def\cM{{\cal M}}

\def\cO{{\cal O}}

\def\cS{{\cal S}}

\def\cV{{\cal V}}
\def\cW{{\cal W}}

\def\mB{\mathfrak{B}}

\def\Im{{\rm Im \,}}

\def\half{{1\over 2}}
\def\p{\partial}

\def\ep{\varepsilon}
\def\om{\omega}

\def\R{{\mathds R}}

\def\C{{\mathds C}}

\def\pbw{\p_{\bar w}}

\def\KK{K}

\def\no{\nonumber}
\def\sm{\smallskip}

\usepackage{bbold}

\usepackage{verbatim} 

\usepackage{titling}
\usepackage{cite}

\numberwithin{equation}{section} 

\begin{document}
\starttext
\setcounter{footnote}{0}

\title{{ $AdS_{2} \times S^{6}$ versus $AdS_{6} \times S^{2}$ in Type IIB supergravity}}
\author{{\bf David Corbino, Eric D'Hoker, Christoph F.~Uhlemann} \\ 
{\sl  Mani L. Bhaumik Institute for Theoretical Physics}\\
{\sl Department of Physics and Astronomy} \\
{\sl University of California, Los Angeles, CA 90095, USA}\\
{\tt \small corbino@physics.ucla.edu, dhoker@physics.ucla.edu, uhlemann@physics.ucla.edu}}

\date{2017 December 11}

\begin{titlingpage}
\maketitle

\begin{abstract}
We obtain the complete local solutions with 16 supersymmetries to Type IIB supergravity on a space-time of the form $AdS_{2}\times S^{6}$ warped over a Riemann surface $\Sigma$ in terms of two locally holmorphic functions on $\Sigma$. We construct the general Ansatz for the bosonic supergravity fields and supersymmetry generators compatible with the $SO(2,1)\oplus SO(7)$ isometry algebra of space-time, which extends to the corresponding real form of the exceptional Lie superalgebra $F(4)$. We  reduce the BPS equations to this Ansatz, obtain their general local solutions, and show that these local solutions solve the full Type IIB supergravity field equations and Bianchi identities. We contrast the $AdS_{2}\times S^{6}$ solution with the closely related $AdS_6\times S^2$ case and present our results for both in parallel. Finally, we present a preliminary analysis of positivity and regularity conditions for $AdS_{2}\times S^{6}$, but postpone the construction of globally regular solutions to a subsequent paper.
\end{abstract}

\end{titlingpage}
\vfill\eject
\setcounter{tocdepth}{1} 
\tableofcontents
\vfill\eject

\baselineskip=15pt
\setcounter{equation}{0}
\setcounter{footnote}{0}

\section{Introduction}

Half-BPS solutions to Type IIB supergravity were obtained recently for a space-time of the form $AdS_{6}\times S^{2}$ warped over a  Riemann surface  \cite{D'Hoker:2016rdq, DHoker:2016ysh, DHoker:2017mds, DHoker:2017zwj}. The motivation for that work was the construction of holographic duals to five-dimensional superconformal field theories (SCFTs). The $SO(2,5)\oplus SO(3)$ isometry algebra of the space-time manifold extends to invariance under the corresponding real form of the exceptional Lie superalgebra $F(4)$, which is the unique superconformal algebra in 5 space-time dimensions. 

\sm

A closely related problem is to construct solutions to Type IIB supergravity whose space-time is of the form $AdS_{2}\times S^{6}$ warped over a Riemann surface $\Sigma$. The isometry algebra  $SO(2,1)\oplus SO(7)$ now extends to a different real form of the exceptional Lie superalgebra $F(4)$, which is one of the superconformal algebras in 2 dimensions with 16 supercharges \cite{VanProeyen:1986me,DHoker:2008wvd}. The relation between these two problems is similar to the one encountered between gravity duals to Wilson loops \cite{D'Hoker:2007fq} and interface solutions \cite{D'Hoker:2007xy}. Experience with such relations through ``double analytic continuation" between supergravity on spaces with $AdS_{p}\times S^{q}$ and $ AdS_{q}\times S^{p}$ factors reveals that their solutions are closely related mathematically, yet their physical space-time structure is quite different.  The existence of half-BPS warped $AdS_{6}\times S^{2}$ solutions to Type~IIB supergravity therefore motivates the search for warped $AdS_{2}\times S^{6}$ solutions.

\sm

The study of $AdS_{2}$ holography provides further motivation for the search for $AdS_{2}\times S^{6}$ supergravity solutions.  Holography on two-dimensional Anti-de Sitter space-time is arguably less well understood than its higher-dimensional counterparts  due in part to certain exotic features of $AdS_2$. They include the presence of multiple disconnected time-like boundary components,  the suppression of finite-energy excitations due to strong gravitational backreaction \cite{Strominger:1998qg, Maldacena:1998fr}, and the ambiguity in identifying the natural CFT dual either as conformal quantum mechanics or as two-dimensional boundary conformal field theory. Much recent focus has been on analyzing the gravitational backreaction on $AdS_{2}$, and on possible relations with dilaton gravity and the Sachdev-Ye-Kitaev model (see  \cite{Maldacena:2016ms} and references therein).

\sm

In this paper, we construct the local form of half-BPS solutions to Type IIB supergravity for warped $AdS_{2}\times S^{6}$ following the strategy used for  $AdS_{6}\times S^{2}$. We derive the reduced BPS equations for the  general Ansatz  dictated by $SO(2,1)\oplus SO(7)$ isometry for the  fields of Type IIB supergravity. These equations are very closely related to the reduced BPS equations for the $AdS_6 \times S^2$ case, but differ by subtle and crucially important signs and factors of $i=\sqrt{-1}$. We provide a detailed comparison between the mathematical equations for both cases.  

\sm

Using methods which are analogous to the ones developed to solve the reduced BPS equations for the $AdS_6 \times S^2$ case, we construct the general local solutions for the $AdS_2 \times S^6$ case in terms of two locally holomorphic functions $\mathcal{A}_{\pm}$ on the Riemann surface $\Sigma$. The differences between the $AdS_{2}\times S^{6}$ and $AdS_6 \times S^2$ solutions are again subtle, but crucial, and to facilitate direct comparisons we discuss both cases in parallel. To solve the reduced BPS equations, we make use of the solution to the axion-dilaton Bianchi identities, but derive the Bianchi identity for the 3-form field strength from the BPS equations. To complete the discussion,  we verify that the full set of Type IIB field equations are satisfied when the bosonic supergravity fields are given by the solutions to the BPS equations and axion-dilaton Bianchi identities. We show this for the $AdS_2\times S^6$ and $AdS_6\times S^2$ cases in parallel, and thus provide this check also for the solutions constructed in \cite{D'Hoker:2016rdq}.

\sm

The solutions obtained for the supergravity fields satisfy the BPS and field equations, but they become physically viable only after certain reality, positivity and regularity conditions are enforced.
We obtain the constraints on the functions $\cA_\pm$ required by physical positivity and regularity conditions on the supergravity fields, and exhibit crucial differences between the $AdS_2\times S^6$ and $AdS_6\times S^2$ cases.  We discuss the possibility of performing a ``double analytic continuation" of the global $AdS_{6}\times S^{2}$ solutions constructed in \cite{DHoker:2016ysh,DHoker:2017mds} to the present case of $AdS_{2}\times S^{6}$. Although such continuations are found to satisfy the field equations, they appear to be neither supersymmetric nor physically regular. Therefore, the construction of global $AdS_2 \times S^6$ solutions must be conducted independently of the $AdS_6 \times S^2$ case. With that objective in mind, we derive the explicit forms of the two-form and six-form potentials for the $AdS_2\times S^6$ and the $AdS_6\times S^2$ solutions. For the $AdS_6\times S^2$ case the two-form potential was a valuable indicator for brane sources in the physically regular solutions constructed in \cite{DHoker:2016ysh,DHoker:2017mds}, and we expect the six-form potential to play a similar role for $AdS_2\times S^6$.

\sm

Solving the physical positivity and regularity conditions, constructing global solutions, obtaining their holographic CFT dual, developing a brane interpretation, and exploring potential relations with $(p,q)$ string webs in analogy to the $(p,q)$ five-brane webs for $AdS_6 \times S^2$ solutions,  are important topics left to subsequent work.

\sm

The remainder of this paper is organized as follows. In section \ref{sec:IIB} we review Type IIB supergravity and introduce the general $SO(2,1) \oplus SO(7)$-invariant Ansatz.  In section \ref{sec:reduce}  we reduce the BPS equations to this Ansatz. In section \ref{sec:local-sol}, we solve the reduced BPS equations in terms of two locally holomorphic functions on $\Sigma$. In section \ref{sec:sugra-sol-summary} we obtain the expressions for the supergravity fields of the solutions in terms of the holomorphic data, compare with  the $AdS_6 \times S^2$ solutions, and analyze their behavior under $SU(1, 1)$ symmetry.  In section \ref{app:EOM} we  verify that the solutions to the BPS equations obey the field equations for both $AdS_{2}\times S^{6}$ and $AdS_{6}\times S^{2}$ cases. In section \ref{sec:regularity} we derive the physical positivity and regularity conditions, as well as discuss their implications for global solutions. In section \ref{sec:doubleWick}, we discuss the relation between $AdS_{6}\times S^{2}$ and $AdS_{2}\times S^{6}$ provided by double analytic continuation. We conclude in section \ref{sec:conclusion}. In appendix \ref{sec:Clifford}, a basis for the Clifford algebra adapted to our Ansatz is presented. Details of the derivation of the BPS equations are given in appendix \ref{sec:BPS}, of the two- and six-form potentials in appendix \ref{sec:flux}, and of the Ricci tensor components in appendix \ref{sec:Ricci}.


\newpage


\section{ \texorpdfstring{$AdS_{2} \times S^{6} \times \Sigma$}{AdS2xS6xSigma} Ansatz in Type IIB supergravity}\label{sec:IIB}

In this section we begin by reviewing the salient features of Type IIB supergravity needed in this paper, and then obtain the $SO(2,1)\oplus SO(7)$-invariant Ansatz for the bosonic supergravity fields and the generator of supersymmetry transformations.

\subsection{Type IIB supergravity review}

The bosonic fields of Type IIB supergravity consist of the metric $g_{MN}$, the complex-valued axion-dilaton field $B$, a complex-valued  two-form potential $C_{(2)}$ and a real-valued four-form field $C_{(4)}$. The field strengths of the potentials $C_{(2)}$ and $C_{(4)}$ are given as follows,
\bea
F_{(3)} & = & d C_{(2)}
\no \\
F_{(5)} & = & dC_{(4)} + \frac{i}{16}(C_{(2)} \wedge \bar{F}_{(3)} - \bar{C}_{(2)} \wedge F_{(3)})
\eea
The field strength $F_{(5)}$ satisfies the well-known self-duality condition $F_{(5)} = *F_{(5)}$. Instead of the scalar field $B$ and the 3-form $F_{(3)}$, the fields that actually enter the BPS equations are composite fields, namely the one-forms $P,Q$ representing $B$, and the complex 3-form $G$ representing $F_{(3)}$, given in terms of the fields defined above by the following relations,
\bea
\label{PQG}
P & = & f^2 \, d B \hskip 1.5in f^2 = ( 1 - |B|^2)^{-1}
\no \\
Q & = & f^2 \, \Im (B d \bar B) 
\no \\
G & = & f (F_{(3)} - B\bar{F}_{(3)})
\eea
Under the $SU(1,1)\sim SL(2,\R)$ global symmetry of Type IIB supergravity, the Einstein-frame metric $g_{MN}$ and the four-form $C_{(4)}$ are invariant, while $B$ and $C_{(2)}$ transform as,
\bea
\label{eq:2_2a} 
B & \to &  (uB + v)/(\bar{v}B + \bar{u}) 
\no \\
C_{(2)} & \to &  uC_{(2)} + v\bar{C}_{(2)}
\eea
where $SU(1,1)$ is parametrized by  $u, v \in \C$ with $|u|^{2} - |v|^{2} = 1$. The field $B$ takes values in the coset $SU(1,1)/U(1)_{q}$ and $Q$ plays the role of a composite $U(1)_q$ gauge field. The transformation laws for the composite fields are as follows \cite{Schwarz:1983qr},
\bea
\label{eq:2_2b} 
P & \to & e^{2i\theta}P \hskip 1.5in \theta = \arg (v \bar B +u)
\no \\
Q & \to & Q + d\theta 
\no \\
G & \to & e^{i\theta}G
\eea
Equivalently, one may formulate Type IIB supergravity directly in terms of $g_{MN}$, $F_{(5)}$, $P,Q$ and $G$
provided these fields are subject to the Bianchi identities \cite{Schwarz:1983qr, Howe:1983sra}.

\sm

The fermionic fields are Weyl fermions with opposite 10-dimensional chirality, namely  the dilatino $\lambda$ satisfying $\Gamma_{11}\lambda = \lambda$ and the gravitino $\psi_{M}$ satisfying $\Gamma_{11}\psi_{M} = -\psi_{M}$. The crucial information for the construction of supersymmetric solutions to Type IIB supergravity are the supersymmetry variations of the fermions, respectively $\delta \lambda$ and $\delta \psi_M$. The BPS equations are the conditions that the fermion fields and their variations vanish, and are given by,\footnote{The signature convention for the metric is $(- + \cdots +)$, the Dirac-Clifford algebra is defined by the relation $\{ \Gamma^{M}, \Gamma^{N} \} = 2\eta^{MN}I_{32}$ and the charge conjugation matrix $\mathcal{B}$ is defined by the relations $\mathcal{B}^{*}\mathcal{B} = I$ and $\mathcal{B}\Gamma^{M}\mathcal{B}^{-1} = (\Gamma^{M})^{*}$. Repeated indices are summed over, as usual, and complex conjugation is denoted by a \textit{bar} for functions and by a \textit{star} for spinors. We will also use the notation $\Gamma\cdot T \equiv \Gamma^{M_{1}\cdots M_{p}}T_{M_{1}\cdots M_{p}}$ for the contraction of an antisymmetric tensor field $T$ of rank $p$ with a $\Gamma$-matrix of the same rank.} 
\bea 
\label{eq:2_1d} 
0 & = & i(\Gamma\cdot P)\mathcal{B}^{-1}\varepsilon^{*} - \frac{i}{24}(\Gamma\cdot G)\varepsilon 
\no \\
0 & = & (\nabla_{M} - { i \over 2} Q_M) \varepsilon + \frac{i}{480}(\Gamma\cdot F_{(5)})\Gamma_{M}\varepsilon - \frac{1}{96}(\Gamma_{M}(\Gamma\cdot G) + 2(\Gamma\cdot G)\Gamma_{M})\mathcal{B}^{-1}\varepsilon^{*}
\eea
Here, $\ep $ is the generator of infinitesimal supersymmetry transformations. It transforms under the minus chirality Weyl spinor representation of $SO(1,9)$ and $\nabla _M$ is the covariant derivative acting on this representation.   In sec.~\ref{app:EOM} we will show that the solutions with 16 supersymmetries to these BPS equations satisfy the field equations, and when formulated in terms of $P,Q,G,F_{(5)}$ also satisfy their Bianchi identities.

\subsection{\texorpdfstring{$SO(2,1) \oplus SO(7)$}{SO(2,1)xSO(7)} invariant Ansatz for supergravity fields}\label{sec:ansatz}

We construct a general Ansatz for the bosonic fields of Type IIB supergravity invariant or covariant under the $SO(2,1) \oplus SO(7)$ symmetry algebra. The $SO(2,1)$ and $SO(7)$ parts are realized by a geometry which contains  a factor $AdS_{2}$ as well as a factor $S^{6}$, so that the space-time is given by,
\begin{equation}\label{eq:2_3b} 
AdS_{2} \times S^{6} 
\end{equation}
warped over a two-dimensional space $\Sigma$. To produce  a geometry of Type IIB supergravity,  $\Sigma$ has to be orientable and carry a Riemannian metric, and is therefore a Riemann surface. The resulting $SO(2,1) \oplus SO(7)$-invariant Ansatz for the metric can be written as,
\begin{equation}\label{eq:2_3c} 
ds^{2} = f^{2}_{2} \, d\hat s^{2}_{AdS_{2}} + f^{2}_{6} \, d\hat s^{2}_{S^{6}} + ds^{2}_{\Sigma}
\end{equation}
where $f_{2}$, $f_{6}$, and $ds^{2}_{\Sigma}$ are functions of $\Sigma$. We  introduce an orthonormal frame,
\begin{align}\label{eq:2_3d} 
e^{m} &= f_{2} \, \hat{e}^{m} && m = 0,1 \nonumber \\
e^{i} &= f_{6}\, \hat{e}^{i} && i =  2, 3, 4, 5, 6, 7 \nonumber \\
e^{a} & && a = 8, 9
\end{align}
where $\hat{e}^{m}$ and $\hat{e}^{i}$ respectively refer to orthonormal frames for the spaces $AdS_{2}$ and $S^{6}$ with unit radius and ${e}^{a}$ is an orthonormal frame on $\Sigma$ only. In particular, we have,
\begin{align}\label{eq:2_3e} 
d\hat s^{2}_{AdS_{2}} &= \eta^{(2)}_{mn}\hat{e}^{m} \otimes \hat{e}^{n} && \eta^{(2)} = \textrm{diag}(- +) \nonumber \\
d\hat s^{2}_{S^{6}} &= \delta_{ij}\hat{e}^{i} \otimes \hat{e}^{j} && \nonumber \\
ds^{2}_{\Sigma} &= \delta_{ab} e^{a} \otimes  e^{b} &&
\end{align}
The requirement for $SO(2,1) \oplus SO(7)$-invariance restricts $F_{(5)}=0$ as well as,
\bea
\label{eq:2_3f} 
P = p_{a}e^{a} 
\hskip 0.7in 
Q = q_{a}e^{a} 
\hskip 0.7in 
 G = g_{a}e^{a} \wedge e^{0} \wedge e^{1}
\eea
where  the components $p_{a}$, $q_{a}$, and $g_{a}$ are complex and depend on $\Sigma$ only.
We have thus parametrized the entire configuration in terms of functions that have non-trivial dependence only on $\Sigma$,
and it will be convenient to set up the frame and coordinates on $\Sigma$ more explicitly.
We will use complex frame indices $z$, $\bar z$ with the following conventions,
\bea
\label{eq:3_4m} 
\delta_{z\bar{z}}  = 2 
\hskip 0.6in  
\delta^{z\bar{z}}  = \frac{1}{2}
\hskip 0.6in  
e^{z}  = \frac{1}{2}(e^{8} + ie^{9}) 
\hskip 0.6in  
e^{\bar{z}}  = \frac{1}{2}(e^{8} - ie^{9}) 
\eea
We introduce local complex coordinates $w$, $\bar w$ such that the metric on $\Sigma$ reads,
\begin{align}
 ds^2_\Sigma&=4\rho^2|dw|^2
\end{align}
and we have,
\begin{align}\label{eq:4_2b} 
e^{z} & = \rho dw & D_{z} & = \rho^{-1}\partial_{w} & \hat{\omega}_{z} & = i\rho^{-2}\partial_{w}\rho \nonumber \\
e^{\bar{z}} & = \rho d\bar{w} & D_{\bar{z}} & = \rho^{-1}\partial_{\bar{w}} & \hat{\omega}_{\bar{z}} & = -i\rho^{-2}\partial_{\bar{w}}\rho
\end{align}
This completes the Ansatz for the bosonic fields.

\subsection{\texorpdfstring{$SO(2,1) \oplus SO(7)$}{SO(2,1)xSO(7)} invariant Ansatz for susy generators}

Next, we decompose the supersymmetry generator spinor $\ep$ in an $SO(2,1) \oplus SO(7)$-invariant basis of Killing spinors.  The Killing spinor equations on $AdS_{2}$ and on $ S^{6}$ were derived in Appendix B of \cite{D'Hoker:2007fq} and are respectively given by,
\begin{align}
\label{eq:3_1a} 
\Big ( \hat{\nabla}_{m} - \frac{1}{2}\eta_{1} \gamma_{m} \otimes I_{8} \Big ) \chi^{\eta_{1},\eta_{2}}_{\alpha} & = 0 \nonumber \\
\Big ( \hat{\nabla}_{i} - \frac{i}{2}\eta_{2} I_{2} \otimes \gamma_{i} \Big ) \chi^{\eta_{1},\eta_{2}}_{\alpha} & = 0
\end{align}
where $m$ and $i$ are all {\sl frame indices}. Note that $\hat{\nabla}_{m}$ and $\hat{\nabla}_{i}$ stand for the covariant spinor derivatives respectively on the spaces $AdS_{2}$ and $S^{6}$ with unit radius. The spinors $\chi^{\eta_{1},\eta_{2}}_{\alpha}$ are 16-dimensional, and the parameters $\eta_{1}$ and $\eta_{2}$ can take the values $\pm $. For each value of $(\eta_{1},\eta_{2})$, these equations admit solutions with a four-fold degeneracy, which is labelled by the index $\alpha = 1, 2, 3, 4$. The action of the chirality matrices is given by,
\begin{align}
\label{eq:3_1b} 
\left( \gamma_{(1)} \otimes I_{8} \right) \chi^{\eta_{1},\eta_{2}}_{\alpha} & = \chi^{-\eta_{1},\eta_{2}}_{\alpha} \nonumber \\
\left( I_{2} \otimes \gamma_{(2)} \right) \chi^{\eta_{1},\eta_{2}}_{\alpha} & = \chi^{\eta_{1},-\eta_{2}}_{\alpha}
\end{align}
These equations can be understood as follows. Beginning with $\eta_{1} = \eta_{2} = +$, we pick a basis $\chi^{+,+}_{\alpha}$ for the four-dimensional vector space of spinors for fixed $\eta_{1}, \eta_{2}$ such that the action of $\gamma_{(1)}$ and $\gamma_{(2)}$ are diagonal. The basis for $\chi^{\eta_{1},\eta_{2}}_{\alpha}$ can then simply be defined for the remaining three values of $\eta_{1}, \eta_{2}$ by the action of the chirality matrices above.

\sm

Using arguments similar to the ones used for the $AdS_6 \times S^2$ case, we relate the complex conjugate basis spinors to the original basis by, 
\begin{equation}\label{eq:3_1d} 
\left( B^{-1}_{(1)}\otimes B^{-1}_{(2)} \right) (\chi^{\eta_{1},\eta_{2}}_{\alpha})^{*} = \eta_{2}\chi^{\eta_{1},\eta_{2}}_{\alpha}
\end{equation}
for all values of $\eta_{1}$, $\eta_{2}$, and $\alpha$. Since this decomposition is now canonical in terms of the degeneracy index $\alpha$, we will no longer indicate it explicitly. An arbitrary 32-component complex spinor $\varepsilon$ may be decomposed onto the above Killing spinors as follows,
\begin{equation}\label{eq:3_2a} 
\varepsilon = \sum_{\eta_{1},\eta_{2} = \pm}\chi^{\eta_{1},\eta_{2}}\otimes\zeta_{\eta_{1},\eta_{2}}
\end{equation}
where $\zeta_{\eta_{1},\eta_{2}}$ is a  complex 2-component spinor for each $\eta_{1}$, $\eta_{2}$, and the four-fold degeneracy index is suppressed. As a supersymmetry generator in Type IIB, the spinor $\varepsilon$ must be of definite chirality $\Gamma^{11}\varepsilon = -\varepsilon$, which imposes the following chirality requirements on $\zeta$,
\begin{equation}\label{eq:3_2b} 
\gamma_{(3)}\zeta_{-\eta_{1},-\eta_{2}} = -\zeta_{\eta_{1},\eta_{2}}
\end{equation}
Finally, the charge conjugate spinor is given by,
\begin{align}\label{eq:3_2d} 
\mathcal{B}^{-1}\varepsilon^{*} & = \sum_{\eta_{1},\eta_{2}}\chi^{\eta_{1},\eta_{2}}\otimes\star\zeta_{\eta_{1},\eta_{2}} & \star\zeta_{\eta_{1},\eta_{2}} & = -i\eta_{2}\sigma^{2}\zeta^{*}_{\eta_{1},-\eta_{2}}
\end{align}
This completes the construction of the $SO(2,1) \oplus SO(7)$-invariant Ansatz.

\newpage

\section{Reducing the BPS equations}\label{sec:reduce}

The residual supersymmetries, if any, of a configuration of purely bosonic Type IIB supergravity fields are governed by the BPS equations of (\ref{eq:2_1d}). As we will discuss in more detail in sec.~\ref{app:EOM}, any $SO(2,1) \oplus SO(7)$ invariant Ansatz for the supergravity fields and for the 16-component supersymmetry spinor as discussed in sec.~\ref{sec:ansatz} which satisfies the BPS equations will   automatically solve the Bianchi and field equations, and thus automatically provides a half-BPS solution to Type IIB supergravity.

\sm

In this section, we reduce the BPS equations to the $AdS_{2} \times S^{6} \times \Sigma$ Ansatz, expose its residual symmetries, and solve those reduced equations which are purely algebraic in the supersymmetry spinor components. This will produce simple algebraic expressions for the metric factors $f_{2}$, $f_{6}$ in terms of the spinors. The remaining reduced BPS equations will  be solved for the remaining bosonic fields in subsequent sections. The strategy employed here is the same as the one used in \cite{D'Hoker:2016rdq}.

\subsection{The reduced BPS equations}

We use the $\tau$ matrix notation introduced originally in \cite{Gomis:2006cu} in order to compactly express the action of the various $\gamma$ matrices on the reduced supersymmetry generator $\zeta$ introduced in (\ref{eq:3_2a}). Defining $\tau^{(ij)} = \tau^{i}\otimes\tau^{j}$ with $i,j = 0, 1, 2, 3$, we identify $\tau^{0}$ with the identity matrix and $\tau^{i}$ for $i = 1, 2, 3$ with the standard Pauli matrices. The action of these matrices on $\zeta$ may be written in components as follows, 
\begin{equation}\label{eq:3_2e} 
( \tau^{(ij)}\zeta )_{\eta_{1},\eta_{2}} \equiv \sum_{\eta_{1}',\eta_{2}'}(\tau^{i})_{\eta_{1}\eta_{1}'}(\tau^{j})_{\eta_{2}\eta_{2}'}\zeta_{\eta_{1}'\eta_{2}'}
\end{equation}
The reduced BPS equations may then be calculated using the decomposition of $\varepsilon$ onto Killing spinors given in (\ref{eq:3_2a}). The reduced dilatino equation is given by,
\begin{equation}\label{eq:3_3a} 
0 = -4p_{a}\gamma^{a}\sigma^{2}\zeta^{*} + g_{a}\tau^{(12)}\gamma^{a}\zeta
\end{equation}
while the reduced gravitino equations take the following form,
\begin{align}\label{eq:3_3b} 
& (m) & 0 & = \frac{-i}{2f_{2}}\tau^{(21)}\zeta + \frac{D_{a}f_{2}}{2f_{2}}\gamma^{a}\zeta - \frac{3}{16}g_{a}\tau^{(12)}\gamma^{a}\sigma^{2}\zeta^{*} \nonumber \\
& (i) & 0 & = \frac{1}{2f_{6}}\tau^{(02)}\zeta + \frac{D_{a}f_{6}}{2f_{6}}\gamma^{a}\zeta + \frac{1}{16}g_{a}\tau^{(12)}\gamma^{a}\sigma^{2}\zeta^{*} \nonumber \\
& (a) & 0 & = \left( D_{a} + \frac{i}{2}\hat{\omega}_{a}\sigma^{3} - \frac{i}{2}q_{a} \right)\zeta - \frac{3}{16}g_{a}\tau^{(12)}\sigma^{2}\zeta^{*} + \frac{1}{16}g_{b}\tau^{(12)}\gamma_{a}{}^{b}\sigma^{2}\zeta^{*}
\end{align}
The derivative $D_{a}$ acts on functions of $\Sigma$ only, and is defined with respect to the frame $e^{a}$ of $\Sigma$, so that the total differential $d_\Sigma$ on $\Sigma$  takes the form $d_\Sigma  = e^{a}D_{a}$, while the $U(1)$-connection with respect to frame indices is $\hat{\omega}_{a}$. The reduction is carried out in Appendix \ref{sec:BPS}.

\subsection{Symmetries of the reduced BPS equations}

The global $SU(1,1)$ symmetry of Type IIB supergravity, whose action on the bosonic fields was given in (\ref{eq:2_2a}) and (\ref{eq:2_2b}), survives the reduction to the $SO(2,1) \oplus SO(7)$ invariant Ansatz. 
It leaves the metric functions $f_2, f_6, \rho$ invariant, transforms the axion-dilaton field $B$ and the two-form $C_{(2)}$ as in (\ref{eq:2_2a}), transforms the reduced supersymmetry spinor $\zeta$ by $\zeta \to e^{i \theta /2} \zeta$,  and transforms the composite fields of (\ref{eq:2_2b}) as follows, 
\begin{align}\label{eq:3_3ba} 
U(1)_q : \, q_{a} \to q_{a} + D_{a}\theta 
\hskip 0.6in 
p_{a} \to e^{2i\theta}p_{a} 
\hskip 0.6in
g_a \to g_{a} e^{i\theta}g_{a}
\end{align}
The reduced BPS equations are also invariant under the following discrete symmetries which act only on the reduced supersymmetry generator but not on the reduced supergravity fields,
\begin{align}\label{eq:3_3c} 
\cI: \, \zeta \to -\tau^{(11)}\sigma^{3}\zeta 
\hskip 1in
\cJ: \, \zeta \to  \tau^{(32)}\zeta
\end{align}
Finally, charge conjugation $\cK$ acts by, 
\begin{align}\label{eq:3_3ca} 
\cK : \, \zeta  \to  \tau^{(22)}\sigma^{2}\zeta^{*} 
\hskip 0.5in
q_{a}  \to  -q_{a}  
\hskip 0.5in
p_{a}  \to  \bar{p}_{a} 
\hskip 0.5in
g_{a}  \to  -\bar{g}_{a}
\end{align}
The chirality requirement of Type IIB restricts the spinor $\zeta$ to the subspace,
\begin{equation}\label{eq:3_3d} 
\mathcal{I}\zeta = -\tau^{(11)}\sigma^{3}\zeta = \zeta
\end{equation}
The symmetries $\mathcal{I}$, $\mathcal{J}$, $\mathcal{K}$ commute with one another and may be diagonalized simultaneously. Both $\mathcal{I}$ and $\mathcal{J}$ commute with $U(1)_{q}$, but $\cK$ does not commute with $U(1)_q$.

\subsection{Restricting to a single subspace of \texorpdfstring{$\mathcal{J}$}{cJ}}

The eigenspace of $\cI$ being already restricted by the chirality condition of (\ref{eq:3_3d}), we now derive the restrictions to the eigenspaces of $\mathcal{J}$ and $\mathcal{K}$ which are implied by the reduced BPS equations, following the same procedure that was used for $AdS_6 \times S^2$ in \cite{D'Hoker:2016rdq}. 
For any value of $g_a$, we have the following quadratic relations in $\zeta$,
\begin{equation}\label{eq:3_4a} 
g_{a}\zeta^{t}T\tau^{(12)}\sigma^{2}\gamma^{a}\zeta = 0
\end{equation}
provided the matrix $T$ belongs to the following set of $\tau^{(ij)}$-matrices, 
\begin{equation}\label{eq:3_4b} 
T \in \mathcal{T} = \left\{ \tau^{(00)}, \tau^{(10)}, \tau^{(20)}, \tau^{(31)}, \tau^{(32)}, \tau^{(33)} \right\}
\end{equation}
For these values of $T$, the combination $T\tau^{(12)}\sigma^{2}\gamma^{a}$ is anti-symmetric for $a=1,2$ and the relation (\ref{eq:3_4a})  indeed holds automatically. The reduced dilatino equation implies,
\begin{equation}\label{eq:3_4c} 
\bar{p}_{a}\zeta^{\dagger}T\gamma^{a}\zeta = 0
\end{equation}
Making use also of the chirality condition (\ref{eq:3_3d}), we obtain the further equation,
\begin{equation}\label{eq:3_4d} 
\bar{p}_{a}\zeta^{\dagger}T\tau^{(11)}\gamma^{a}\sigma^{3}\zeta = 0
\end{equation}
When both $T$ and $T\tau^{(11)}$ belong to $\mathcal{T}$, which is the case for only a single pair of matrices, namely $T = \tau^{(20)} $ or $T = \tau^{(31)}$, and assuming that $p_a$ does not vanish identically, we may combine (\ref{eq:3_4c}) and (\ref{eq:3_4d}) to obtain the following  relations,
\begin{equation}\label{eq:3_4e} 
\zeta^{\dagger}\tau^{(20)}\gamma^{a}\zeta = \zeta^{\dagger}\tau^{(31)}\gamma^{a}\zeta = 0 
\end{equation}
which hold for $a=1,2$ and  are equivalent to one another upon using the chirality condition.

\sm

Next, we analyze the gravitino equations. Multiplying equations $(m)$ and $(i)$ of (\ref{eq:3_3b}) on the left by $\zeta^{\dagger}T\sigma^{p}$ for $p = 0, 3$, we obtain a cancellation of the last term when $T\tau^{(12)}$ is anti-symmetric (which is the same condition we had for the dilatino equation),
\begin{align}\label{eq:3_4f} 
0 & = -\frac{i}{2f_{2}}\zeta^{\dagger}T\tau^{(21)}\sigma^{p}\zeta + \frac{D_{a}f_{2}}{2f_{2}}\zeta^{\dagger}T\sigma^{p}\gamma^{a}\zeta \nonumber \\
0 & = \frac{1}{2f_{6}}\zeta^{\dagger}T\tau^{(02)}\sigma^{p}\zeta + \frac{D_{a}f_{6}}{2f_{6}}\zeta^{\dagger}T\sigma^{p}\gamma^{a}\zeta
\end{align}
In view of (\ref{eq:3_4e}), the second term will cancel when $T = \tau^{(20)}$ and $T = \tau^{(31)}$. so that we obtain the following relations from the remaining cancellation of the first term,
\begin{align}\label{eq:3_4g} 
\zeta^{\dagger}\tau^{(01)}\sigma^{p}\zeta & = 0 & p & = 0, 3 \nonumber \\
\zeta^{\dagger}\tau^{(22)}\sigma^{p}\zeta & = 0 & &
\end{align}
and their chiral conjugates, which may be obtained by using the chirality condition. 

\sm

Next, we use the general result of \cite{D'Hoker:2007xy} that the bilinear equation $\zeta^{\dagger}M\zeta = 0$ is solved by projecting $\zeta$ onto a subspace via a projection matrix $\Pi$ that anti-commutes with $M$.  Thus, we must find a projector $\Pi$  with the following properties,
\begin{equation}\label{eq:3_4i} 
[\Pi, \tau^{(11)}\sigma^{3}] = \{ \Pi \tau^{(01)}\sigma^{p} \} = \{ \Pi, \tau^{(22)}\sigma^{p} \} = 0
\end{equation}
The solutions to these equations are $\tau^{(32)}$, $\tau^{(23)}$, $\tau^{(32)}\sigma ^3 $, and $\tau^{(23)} \sigma ^3$. These four possibilities are pairwise equivalent under the chirality relation. The projector $\Pi = \tau^{(32)}$ precisely corresponds to the symmetry $\mathcal{J}$, so imposing a restriction on the spinor space by this operator is the only consistent restriction. Therefore, we will impose the restriction,
\begin{align}
\label{eq:3_4j} 
\tau^{(32)}\zeta & = \nu\zeta & \nu & = \pm 1
\end{align}
which solves all the above bilinear relations for either choice of $\nu$, but not both. 

\sm

We may solve the projection relations given in (\ref{eq:3_3d}) and  (\ref{eq:3_4j})  in terms of  two independent complex-valued one-component spinors $\alpha$ and $\beta$. Denoting the components of $\zeta$ by $\zeta_{abc}$, where  $a, b$  label the $\tau$-matrix basis, while $c$ labels the chirality basis in which $\sigma^{3}$ is diagonal, and $a,b,c$  take values $\pm$, we have, 
\begin{align}\label{eq:3_4l} 
\bar{\alpha} & = \zeta_{+++} = -\zeta_{--+} = -i\nu\zeta_{+-+} = +i\nu\zeta_{-++} \nonumber \\
\beta & = \zeta_{---} = +\zeta_{++-} = -i\nu\zeta_{-+-} = -i\nu\zeta_{+--}
\end{align}

\subsection{The reduced BPS equations in component form}

To decompose the reduced BPS equations in a basis of complex frame indices $z$, $\bar{z}$, we use (\ref{eq:3_4m}) along with the following basis of $\gamma$-matrices compatible with a diagonal $\sigma ^3$, 
\begin{align}\label{eq:3_4m2}
\gamma^{z} & = \frac{1}{2}(\gamma^{8} + i\gamma^{9}) = \begin{pmatrix} 
0 & 1\\ 
0 & 0 
\end{pmatrix}
&
\gamma^{\bar{z}} & = \frac{1}{2}(\gamma^{8} - i\gamma^{9}) = \begin{pmatrix} 
0 & 0\\ 
1 & 0 
\end{pmatrix}
\end{align}
Using   (\ref{eq:3_4l})  the reduced dilatino equations become,
\begin{align}\label{eq:3_5a} 
4p_{z}\alpha + g_{z}\beta & = 0 \nonumber \\
4p_{\bar{z}}\bar{\beta} + g_{\bar{z}}\bar{\alpha} & = 0
\end{align}
The gravitino equations which are purely algebraic in $\alpha, \beta, \bar \alpha, \bar \beta$ are given by,
\begin{align}\label{eq:3_5b} 
\frac{1}{2f_{2}}\bar{\alpha} + \frac{D_{z}f_{2}}{2f_{2}}\beta + \frac{3}{16}g_{z}\alpha & = 0 \nonumber \\ -\frac{1}{2f_{2}}\beta + \frac{D_{\bar{z}}f_{2}}{2f_{2}}\bar{\alpha} + \frac{3}{16}g_{\bar{z}}\bar{\beta} & = 0 \nonumber \\
\frac{\nu}{2f_{6}}\bar{\alpha} + \frac{D_{z}f_{6}}{2f_{6}}\beta - \frac{1}{16}g_{z}\alpha & = 0 \nonumber \\
\frac{\nu}{2f_{6}}\beta + \frac{D_{\bar{z}}f_{6}}{2f_{6}}\bar{\alpha} - \frac{1}{16}g_{\bar{z}}\bar{\beta} & = 0 
\end{align}
while the gravitino equations which are differential in $\alpha, \beta, \bar \alpha, \bar \beta$ are given by,
\begin{align}\label{eq:3_5c} 
\left( D_{z} + \frac{i}{2}\hat{\omega}_{z} - \frac{i}{2}q_{z} \right)\bar{\alpha} + \frac{1}{4}g_{z}\bar{\beta} & = 0 \nonumber \\
\left( D_{z} - \frac{i}{2}\hat{\omega}_{z} - \frac{i}{2}q_{z} \right)\beta + \frac{1}{8}g_{z}\alpha & = 0 \nonumber \\
\left( D_{\bar{z}} + \frac{i}{2}\hat{\omega}_{\bar{z}} - \frac{i}{2}q_{\bar{z}} \right)\bar{\alpha} + \frac{1}{8}g_{\bar{z}}\bar{\beta} & = 0 \nonumber \\
\left( D_{\bar{z}} - \frac{i}{2}\hat{\omega}_{\bar{z}} - \frac{i}{2}q_{\bar{z}} \right)\beta + \frac{1}{4}g_{\bar{z}}\alpha & = 0
\end{align}
In addition, we have the complex conjugate equations to all of the equations above. Note that since $G$ and $P$ are complex-valued, we have in general $(g_{z})^{*} \neq g_{\bar{z}}$ and $(p_{z})^{*} \neq p_{\bar{z}}$.

\subsection{Determining the radii \texorpdfstring{$f_{2}$, $f_{6}$}{f2, f6}}

One may solve for the radii starting from the algebraic gravitino equations of (\ref{eq:3_5b}). Taking the linear combination of the first  equation in (\ref{eq:3_5b}) with coefficient  $\bar{\beta}$ and the complex conjugate of the second equation $\alpha$ in (\ref{eq:3_5b}) with coefficient  $\bar{\alpha}$ on the one hand, and the third equation in (\ref{eq:3_5b}) with coefficient  $- \bar{\beta}$ and the complex conjugate of the fourth equation with coefficient $\bar \alpha$, we obtain the following equations, 
\bea
\label{eq:3_6b} 
\frac{D_{z}f_{2}}{2f_{2}}(\alpha\bar{\alpha} + \beta\bar{\beta}) 
& = & -\frac{3}{16}g_{z}\alpha\bar{\beta} - \frac{3}{16}(g_{\bar{z}})^{*} \bar{\alpha}\beta
\no \\
\frac{D_{z}f_{6}}{2f_{6}}(\alpha\bar{\alpha} - \beta\bar{\beta}) 
& = & -\frac{1}{16}g_{z}\alpha\bar{\beta} + \frac{1}{16}(g_{\bar{z}})^{*} \bar{\alpha}\beta
\eea
These combinations suggest that we should evaluate the covariant derivatives $D_{z}(\alpha\bar{\alpha} \pm \beta\bar{\beta})$ out of the differential equations (\ref{eq:3_5c}) for $\alpha$, $\beta$, $\bar{\alpha}$, $\bar{\beta}$, and we find,
\begin{align}\label{eq:3_6e} 
D_{z}(\alpha\bar{\alpha} + \beta\bar{\beta}) & = -\frac{3}{8}g_{z}\alpha\bar{\beta} - \frac{3}{8}(g_{\bar{z}})^{*} \bar{\alpha}\beta \nonumber \\
D_{z}(\alpha\bar{\alpha} - \beta\bar{\beta}) & = -\frac{1}{8}g_{z}\alpha\bar{\beta} + \frac{1}{8}(g_{\bar{z}})^{*} \bar{\alpha}\beta
\end{align}
Eliminating all flux dependences between (\ref{eq:3_6b}) and (\ref{eq:3_6e}), we may integrate the resulting relations, to obtain the following expressions for the radii, 
\begin{align}\label{eq:3_6g} 
f_{2} & = c_{2}(\alpha\bar{\alpha} + \beta\bar{\beta}) \nonumber \\
f_{6} & = c_{6}(\alpha\bar{\alpha} - \beta\bar{\beta})
\end{align}
where $c_{2}$ and  $c_{6}$ are integration constants.

\subsection{Solving the remaining algebraic gravitino equations}

To obtain the results of the previous subsection, we have taken only pairwise linear combinations of the algebraic gravitino equations. Here, we take the orthogonally conjugate pairwise linear combinations. To guarantee that the four resulting bilinear equations are equivalent to the original four algebraic gravitino equations, we must have that the determinant of the two linear combinations is $\alpha\bar{\alpha} + \beta\bar{\beta} \neq 0$. Therefore, we multiply the first equation by $\alpha$ and the second by $-\beta$, so that the terms in $D_{z}f_{2}$ and $D_{z}f_{6}$ cancel out, and we are left with,
\begin{align}
\label{eq:3_7a} 
\frac{1}{2c_{2}} + \frac{3}{16}g_{z}\alpha^{2} - \frac{3}{16}(g_{\bar{z}})^{*} \beta^{2} & = 0 \nonumber \\
\frac{\nu}{2c_{6}} - \frac{1}{16}g_{z}\alpha^{2} + \frac{1}{16}(g_{\bar{z}})^{*} \beta^{2} & = 0
\end{align}
The last equation may be simplified with the help of the first and yields,
\begin{equation}\label{eq:3_7b} 
c_{6} = -3\nu c_{2}
\end{equation}
Recall that $\nu$ is allowed to take either value $\nu = \pm 1$, but not both.

\subsection{Summary and comparison to \texorpdfstring{$AdS_{6}\times S^{2}$}{AdS6xS2}}\label{sec:eq-summary}

In this subsection, we summarize the remaining reduced BPS equations for the $AdS_2 \times S^6$ case and present the result in parallel with the corresponding results for the remaining reduced BPS equations obtained for the case $AdS_{6}\times S^{2}$  in \cite{D'Hoker:2016rdq}. To this end we introduce the quantities $\KK$ and $c$ to distinguish between the two cases as follows,
\begin{align}\label{eq:LambdaKC}
 K &=i  &  c&=\nu c_6  &  \text{for $AdS_2\times S^6$}
\nonumber\\
 K&=1 &  c&=c_6  &  \text{for $AdS_6\times S^2$}
\end{align}
With the help of these quantities the remaining reduced BPS equations take on a remarkably unified form.
The remaining reduced dilatino equations are,
\begin{align}\label{eq:3_8a}
 -4iK p_z \alpha+g_z\beta&=0
 \nonumber\\
 4 p_{\bar z}\bar\beta-iK g_{\bar z}\bar\alpha&=0
\end{align}
along with their complex conjugates. The radii in terms of the spinors $\alpha, \beta$ are given by,
\begin{align}\label{eq:3_8b} 
f_{2} & = -\frac{\nu}{3}c_{6}(\alpha\bar{\alpha} - \KK^2 \beta\bar{\beta}) \nonumber \\
f_{6} & = c_{6}(\alpha\bar{\alpha} +\KK^2 \beta\bar{\beta})
\end{align}
The remaining algebraic relation between the spinors and the fluxes is given by,
\begin{equation}\label{eq:3_8c} 
\frac{K}{2c} - \frac{i}{16}g_{z}\alpha^{2} + \frac{i}{16}(g_{\bar{z}})^{*} \beta^{2} = 0
\end{equation}
along with its complex conjugate. The remaining differential equations on the spinors are,
\begin{align}
\label{eq:3_8d} 
\left( D_{z} - \frac{i}{2}\hat{\omega}_{z} + \frac{i}{2}q_{z} \right)\alpha + \frac{i}{8K}(g_{\bar{z}})^{*}\beta & = 0 \nonumber \\
\left( D_{z} - \frac{i}{2}\hat{\omega}_{z} - \frac{i}{2}q_{z} \right)\beta + \frac{i}{8K}g_{z}\alpha & = 0 \nonumber \\
\left( D_{z} + \frac{i}{2}\hat{\omega}_{z} - \frac{i}{2}q_{z} \right)\bar{\alpha} - \frac{iK}{4}g_{z}\bar{\beta} & = 0 \nonumber \\
\left( D_{z} + \frac{i}{2}\hat{\omega}_{z} + \frac{i}{2}q_{z} \right)\bar{\beta} - \frac{iK}{4}(g_{\bar{z}})^{*}\bar{\alpha} & = 0
\end{align}
along with their complex conjugates.

\newpage

\section{Local solutions to the BPS equations}\label{sec:local-sol}

The BPS equations have been reduced to the $AdS_{2}\times S^{6}\times\Sigma$ Ansatz and solved for the radii $f_{2}$ and $f_{6}$ in the previous section. In this section, the remaining equations, namely 
the dilatino BPS equations of (\ref{eq:3_8a}), the remaining algebraic gravitino equation of (\ref{eq:3_8c}) and the four differential equations (\ref{eq:3_8d}), will be completely solved locally on $\Sigma$ in terms of two locally holomorphic functions $\cA_\pm$ on $\Sigma$. Thus we will obtain expressions for all bosonic supergravity fields that satisfy the  BPS equations  in terms of $\mathcal{A}_{\pm}$. 

\subsection{Eliminating the reduced flux fields}

We start from the equations summarized in sec.~\ref{sec:eq-summary} and  keep  $K$ and $c$ as defined in (\ref{eq:LambdaKC}) for easier comparison with the $AdS_6\times S^2$ case. We begin by  eliminating the reduced flux fields $g_{z}$, $g_{\bar{z}}$ and their complex conjugates in favor of $p_{z}$, $p_{\bar{z}}$ and their complex conjugates using the dilatino BPS equations of (\ref{eq:3_8a}).
The algebraic relation (\ref{eq:3_8c}) becomes,
\begin{equation}
\label{eq:4_1b} 
p_{z}\frac{\alpha^{3}}{\beta} - (p_{\bar{z}})^{*}\frac{\beta^{3}}{\alpha} + \frac{2}{c} = 0
\end{equation}
The differential equations (\ref{eq:3_8d}) take the following form,
\begin{align}\label{eq:4_1c} 
\left( D_{z} - \frac{i}{2}\hat{\omega}_{z} + \frac{i}{2}q_{z} \right)\alpha - \frac{1}{2}(p_{\bar{z}})^{*}\frac{\beta^{2}}{\alpha} & = 0 \nonumber \\
\left( D_{z} - \frac{i}{2}\hat{\omega}_{z} - \frac{i}{2}q_{z} \right)\beta - \frac{1}{2}p_{z}\frac{\alpha^{2}}{\beta} & = 0 \nonumber \\
\left( D_{z} + \frac{i}{2}\hat{\omega}_{z} - \frac{i}{2}q_{z} \right)\bar{\alpha} + \KK^2 p_{z}\frac{\alpha\bar{\beta}}{\beta} & = 0 \nonumber \\
\left( D_{z} + \frac{i}{2}\hat{\omega}_{z} + \frac{i}{2}q_{z} \right)\bar{\beta} + \KK^2 (p_{\bar{z}})^{*}\frac{\bar{\alpha}\beta}{\alpha} & = 0
\end{align}
Equations (\ref{eq:4_1b}) and (\ref{eq:4_1c}) are the remaining relations to be solved. Their solution will give $\alpha, \beta, f_2, f_6, \rho, p_z$ and therefore $B$ as well as the flux field $g_z, g_{\bar z}$ and their complex conjugates via the reduced dilatino equations.

\subsection{Integrating the first pair of differential equations}

Next, we use the expressions for $p_z, p_{\bar z}, q_z$ and their complex conjugates in terms of the axion-dilation field $B$ via the relations (\ref{PQG}) and (\ref{eq:2_3f}) to solve  the first two equations of (\ref{eq:4_1c})  in terms of holomorphic functions. To do so, we multiply the first equation of (\ref{eq:4_1c}) by $\alpha$ and the second equation of (\ref{eq:4_1c}) by $\beta$,  switch to conformally flat complex coordinates $(w,\bar{w})$ on $\Sigma$ as introduced in (\ref{eq:4_2b}), and use (\ref{PQG}) and (\ref{eq:2_3f}) to express 
$p_{z}$ and $q_{z}$ in terms of $B$,
\begin{align}\label{eq:4_2c} 
\partial_{w}(\rho\alpha^{2}) & = -\frac{1}{2}f^{2}(B\partial_{w}\bar{B} - \bar{B}\partial_{w}B)\rho\alpha^{2} + f^{2}(\partial_{w}\bar{B})\rho\beta^{2} \nonumber \\
\partial_{w}(\rho\beta^{2}) & = +\frac{1}{2}f^{2}(B\partial_{w}\bar{B} - \bar{B}\partial_{w}B)\rho\beta^{2} + f^{2}(\partial_{w}B)\rho\alpha^{2}
\end{align}
By taking suitable linear combinations we obtain the following equivalent equations,
\begin{align}\label{eq:4_2d} 
\partial_{w} \left( \ln\{ \rho(\alpha^{2} - \bar{B}\beta^{2}) \} + \ln f \right) & = 0 \nonumber \\
\partial_{w} \left( \ln\{ \rho(B\alpha^{2} - \beta^{2}) \} + \ln f \right) & = 0
\end{align}
These equations are solved in terms of two independent holomorphic 1-forms $\kappa_{\pm}$, as follows,
\begin{align}\label{eq:4_2e} 
\rho f \left( \alpha^{2} - \bar{B}\beta^{2} \right) & = \bar{\kappa}_{+} \nonumber \\
\rho f \left( \beta^{2} - B\alpha^{2} \right) & = \bar{\kappa}_{-}
\end{align}
Inverting  (\ref{eq:4_2e}), we obtain the spinor components $\alpha$, $\beta$, and their complex conjugates $\bar{\alpha}$, $\bar{\beta}$,
\begin{align}\label{eq:4_2f} 
\rho\alpha^{2} & = f (\bar{\kappa}_{+} + \bar{B}\bar{\kappa}_{-}) & \rho\bar{\alpha}^{2} & = f (\kappa_{+} + B\kappa_{-}) \nonumber \\
\rho\beta^{2} & = f (B\bar{\kappa}_{+} + \bar{\kappa}_{-}) & \rho\bar{\beta}^{2} & = f (\bar{B}\kappa_{+} + \kappa_{-})
\end{align}
The right side of all four equations involves only the holomorphic data $\kappa_{\pm}$ and the $B$-field and their complex conjugates. It remains to solve for the fields $\rho$ and $B$.

\subsection{Preparing the second pair of differential equations}\label{sec:4_3}

Next,  we express the third and fourth equation of (\ref{eq:4_1c}) in terms of $B$, $\rho$ and the local complex coordinates $(w,\bar w)$, and obtain the following set of equations,
\begin{align}
\label{eq:4_3a} 
\left( \partial_{w} - \partial_{w} \ln \rho^2 - f^{2}B\partial_{w}\bar{B} \right)(f\rho\bar{\alpha}^{2}) + 2\KK^2 f^{2}(\partial_{w}B)f\rho\frac{\alpha\bar{\alpha}\bar{\beta}}{\beta} & = 0 \nonumber \\
\left( \partial_{w} -\partial_{w} \ln \rho^2 - f^{2}\bar{B}\partial_{w}B \right)(f\rho\bar{\beta}^{2}) + 2\KK^2 f^{2}(\partial_{w}\bar{B})f\rho\frac{\bar{\alpha}\beta\bar{\beta}}{\alpha} & = 0
\end{align}
The spinors $\alpha, \beta$ and their complex conjugates, as well as the derivatives of $f \rho \bar{\alpha}^2$ and $f \rho \bar{\beta}^2$ may be evaluated in terms of  $B$, $\rho$, and $\kappa _\pm$ and their first derivatives using (\ref{eq:4_1c}). After some simplifications, we obtain the following equivalent system of equations,
\begin{align}
\label{eq:4_3c0} 
\partial_{w}\ln\rho^2 - f^{2}(\partial_{w}\bar{B})\frac{\kappa_{+} + B\kappa_{-}}{\bar{B}\kappa_{+} + \kappa_{-}} - 2\KK^2 f^{2}(\partial_{w}\bar{B})e^{+i\vartheta} & = \frac{\bar{B}\partial_{w}\kappa_{+} + \partial_{w}\kappa_{-}}{\bar{B}\kappa_{+} + \kappa_{-}} \nonumber \\
\partial_{w}\ln\rho^2 - f^{2}(\partial_{w}B)\frac{\bar{B}\kappa_{+} + \kappa_{-}}{\kappa_{+} + B\kappa_{-}} - 2\KK^2 f^{2}(\partial_{w}B)e^{-i\vartheta} & = \frac{\partial_{w}\kappa_{+} + B\partial_{w}\kappa_{-}}{\kappa_{+} + B\kappa_{-}}
\end{align}
where we have used the following abbreviation for the phase angle $\vartheta$,
\begin{align}
\label{eq:4_3d} 
e^{i\vartheta} = \frac{\bar{\alpha}\beta}{\alpha\bar{\beta}}  = \frac{(\kappa_{+} + B\kappa_{-})|B\bar{\kappa}_{+} + \bar{\kappa}_{-}|}{|\bar{\kappa}_{+} + \bar{B}\bar{\kappa}_{-}|(\bar{B}\kappa_{+} + \kappa_{-})} 
\end{align}
The dependence of the algebraic relation (\ref{eq:4_1b}) on $\alpha$ and $\beta$ may also be eliminated using (\ref{eq:4_2f}), while $p_z, p_{\bar z}$ may be expressed in terms of $B$, and we find the equivalent relation,
\begin{equation}
\label{eq:4_3j} 
f^{3}(\partial_{w}B)\frac{(\bar{\kappa}_{+} + \bar{B}\bar{\kappa}_{-})^{\tfrac{3}{2}}}{(B\bar{\kappa}_{+} + \bar{\kappa}_{-})^{\tfrac{1}{2}}} - f^{3}(\partial_{w}\bar{B})\frac{(B\bar{\kappa}_{+} + \bar{\kappa}_{-})^{\tfrac{3}{2}}}{(\bar{\kappa}_{+} + \bar{B}\bar{\kappa}_{-})^{\tfrac{1}{2}}} + \frac{2\rho^{2}}{c} = 0
\end{equation}
Equations (\ref{eq:4_3c0}) and (\ref{eq:4_3j}) are supplemented by their complex conjugates.

\sm

Although on the face of it the remaining equations (\ref{eq:4_3c0}) and (\ref{eq:4_3j}) depend on both $\kappa _\pm$ and their complex conjugates, the conformal invariance of these equations tells us that the dependence 
is actually only through the combination $\rho^2/\kappa _- $ and   the ratio,
\begin{equation}\label{eq:4_3f} 
\lambda = \frac{\kappa_{+}}{\kappa_{-}}
\end{equation}
In terms of these variables, the equations take the following form,
\begin{align}
\label{eq:4_3c} 
\partial_{w}\ln {\rho^2 \over \kappa_-} - f^{2}(\partial_{w}\bar{B})\frac{\lambda + B}{\bar{B}\lambda + 1} - 2\KK^2 f^{2}(\partial_{w}\bar{B})e^{+i\vartheta} 
& = \frac{\bar{B}\partial_{w}\lambda }{\bar{B}\lambda + 1} 
\nonumber \\
\partial_{w}\ln { \rho^2 \over \kappa _-}  - f^{2}(\partial_{w}B)\frac{\bar{B}\lambda + 1}{\lambda + B} - 2\KK^2 f^{2}(\partial_{w}B)e^{-i\vartheta} 
& = \frac{\partial_{w}\lambda }{\lambda + B}
\end{align}
where we now have,
\bea
\label{eq:4_3g} 
e^{i\vartheta} =  \frac{(\lambda + B)|B\bar \lambda + 1|}{|\bar \lambda + \bar{B}|(\bar{B}\lambda + 1)} 
\eea
as well as the  equation resulting from (\ref{eq:4_3j}), 
\begin{equation}\label{eq:4_3k} 
\frac{(\bar{\lambda} + \bar{B})^{\tfrac{3}{2}}}{(B\bar{\lambda} + 1)^{\tfrac{1}{2}}} \, f^3 \partial_{w}B
- \frac{(B\bar{\lambda} + 1)^{\tfrac{3}{2}}}{(\bar{\lambda} + \bar{B})^{\tfrac{1}{2}}} \, f^3 \partial_{w}\bar{B}
+ \frac{2\rho^{2}}{c \, \bar{\kappa}_{-}} = 0
\end{equation}
In summary, we have prepared the remaining reduced BPS equations in the form of complex differential equations (\ref{eq:4_3c}) and (\ref{eq:4_3k}) along with their complex conjugate equations.

\subsection{Decoupling by changing variables}

In this subsection, we will perform two consecutive changes of variables to decouple the remaining equations. 

\subsubsection{First change of variables, from \texorpdfstring{$B$ to $Z$}{B to Z}}

A first change of variables replaces $B$ by a complex field $Z$ and is designed to parametrize the phase $e^{i\vartheta}$ in (\ref{eq:4_3g}) without the square root required from its definition. We make the following rational change of variables to eliminate $B$ in terms of a complex function $Z$,
\begin{align}
\label{eq:4_4_1a} 
Z^{2} & = \frac{\lambda + B}{B\bar{\lambda} + 1} & B & = \frac{Z^{2} - \lambda}{1 - \bar{\lambda}Z^{2}}
\end{align}
which will allow us to express $e^{i\vartheta}$ and $f^{2}$ as rational functions of $Z$ and its complex conjugate,
\begin{align}
\label{eq:4_4_1b} 
e^{i\vartheta} &  = \frac{Z}{\bar{Z}} \left( \frac{1 - \lambda\bar{Z}^{2}}{1 - \bar{\lambda}Z^{2}} \right) & f^{2} & = \frac{(1 - \lambda\bar{Z}^{2})(1 - \bar{\lambda}Z^{2})}{(1 - |\lambda|^{2})(1 - |Z|^{4})}
\end{align}
The equations (\ref{eq:4_3c}) now take the form,
\begin{align}
\partial_{w}\ln {\rho^2 \over \kappa _-} 
& =  
\frac{2Z^{2}\bar{Z} + 4\KK^2 Z}{1 - |Z|^{4}}  \partial_{w}\bar{Z}
 + \frac{\bar{Z}^{2} - \bar{\lambda} + 2\KK^2 Z\bar{Z}^{3} - 2\KK^2 Z\bar{Z}\bar{\lambda}}{(1 - |\lambda|^{2})(1 - |Z|^{4})} \partial_{w}\lambda
\no \\
\partial_{w}\ln {\rho^2 \over \kappa _-} 
& =  
\frac{2 Z^{-1} + 4\KK^2 \bar{Z}}{1 - |Z|^{4}} \partial_{w}Z
 + \frac{Z^{2}\bar{Z}^{2}\bar{\lambda} - \bar{Z}^{2} - 2\KK^2 \bar{Z} Z^{-1} + 2\KK^2 Z\bar{Z}\bar{\lambda} }{(1 - |\lambda|^{2})(1 - |Z|^{4})} \partial_{w}\lambda
\end{align}
Taking the difference of these two equations eliminates the dependence on $\rho^2/\kappa _-$, 
\begin{align}
\label{eq:4_4_1d} 
& (2 + 4\KK^2 |Z|^{2})\partial_{w}Z - Z^{2}(4\KK^2 + 2|Z|^{2})\partial_{w}\bar{Z} 
\nonumber \\
& \qquad = \frac{2\bar{Z}(\KK^2 + |Z|^{2} + \KK^2 |Z|^{4}) - \bar{\lambda}Z(1 + 4\KK^2 |Z|^{2} + |Z|^{4})}{1 - |\lambda|^{2}}\partial_{w}\lambda
\end{align}
while taking their sum gives,
\begin{equation}
\label{eq:4_4_1e} 
\partial_{w}\ln \hat\rho^2 = \frac{1}{2}\partial_{w}\ln\frac{Z}{\bar{Z}} - \KK^2 \frac{\bar{Z}}{Z}\left( \frac{\partial_{w}\lambda}{1 - |\lambda|^{2}} \right)
\end{equation}
where we have changed variables from $\rho$ to $\hat{\rho}$ in the following way,
\begin{equation}
\label{eq:4_4_2b} 
\hat{\rho}^{2} = \frac{\rho^{2}}{c\, \kappa_{-}\bar{\kappa}_{-}}\frac{|1 - Z^{2}\bar{\lambda}|(1 - \KK^2 |Z|^{2})}{f|Z|(1 - |\lambda|^{2})(1 + \KK^2 |Z|^{2})}
\end{equation}
Finally, eliminating $B$ in favor of $Z$ in the algebraic flux equation (\ref{eq:4_3k}) as well, we obtain,
\begin{equation}\label{eq:4_4_1f} 
(1 - |\lambda|^{2})\partial_{w}\left( \frac{Z^{2} + \bar{Z}^{-2}}{1 - |\lambda|^{2}} \right) - \frac{2\partial_{w}\lambda}{1 - |\lambda|^{2}} + 2\hat{\rho}^{2}\kappa_{-}\frac{|Z|}{\bar{Z}^{3}}(1 + \KK^2 |Z|^{2})^{2} = 0
\end{equation}
It remains to solve the system of equations (\ref{eq:4_4_1d}), (\ref{eq:4_4_1e}), and (\ref{eq:4_4_1f}).

\subsubsection{Second change of variables, from \texorpdfstring{$Z$ to $R$, $\psi$}{Z to R, psi}}

A second change of variables is inspired by the form of equation (\ref{eq:4_4_1e}), in which the norm of $Z$ and its phase enter in distinct parts of the equation. We express the complex field $Z$ in terms of two real variables, its absolute value $R$ and phase $\psi$, as follows,
\begin{equation}\label{eq:4_4_2a} 
Z^{2} = R \, e^{i\psi}
\end{equation}
In terms of these variables (\ref{eq:4_4_1e}) takes the form,
\begin{equation}\label{eq:4_4_2c} 
\partial_{w}\ln \hat{\rho}^{2} - \frac{i}{2}\partial_{w}\psi + \KK^2 e^{-i\psi}\frac{\partial_{w}\lambda}{1 - |\lambda|^{2}} = 0
\end{equation}
while (\ref{eq:4_4_1d}) becomes,
\begin{equation}
\label{eq:4_4_2d} 
(1 - R^{2})\frac{\partial_{w}R}{R} + (1 + 4\KK^2 R + R^{2})\left( i\partial_{w}\psi + \frac{\bar{\lambda}\partial_{w}\lambda}{1 - |\lambda|^{2}} \right) - \frac{2\KK^2 e^{-i\psi}(1 + \KK^2 R + R^{2})}{1 - |\lambda|^{2}}\partial_{w}\lambda = 0
\end{equation}
and (\ref{eq:4_4_1f}) becomes,
\begin{equation}
\label{eq:4_4_2e} 
(R^{2} - 1)\frac{\partial_{w}R}{R} + (R^{2} + 1)\left( i\partial_{w}\psi + \frac{\bar{\lambda}\partial_{w}\lambda}{1 - |\lambda|^{2}} \right) - \frac{2R\partial_{w}\lambda}{1 - |\lambda|^{2}}e^{-i\psi} + 2\hat{\rho}^{2}\kappa_{-}e^{i\psi/2}(1 + \KK^2 R)^{2} = 0
\end{equation}
The three equations (\ref{eq:4_4_2c}), (\ref{eq:4_4_2d}), and (\ref{eq:4_4_2e}) are the basic starting point for the complete solution of the full system of reduced BPS equations.

\subsubsection{Decoupling the equations for \texorpdfstring{$\psi$ and $\hat{\rho}^{2}$}{psi and rhohat2}}

Adding equations (\ref{eq:4_4_2d}) and (\ref{eq:4_4_2e}) cancels the terms proportional to $\partial_{w}R$, and concentrates the entire $R$-dependence of this sum in an overall multiplicative factor of $(1 + \KK^2 R)^{2}$. Omitting this factor, the sum becomes,
\begin{equation}
\label{eq:4_4_3a} 
2i\partial_{w}\psi + \frac{2\bar{\lambda}\partial_{w}\lambda}{1 - |\lambda|^{2}} - 2\KK^2 e^{-i\psi} \frac{ \partial_{w}\lambda}{1 - |\lambda|^{2}}  + 2\hat{\rho}^{2}\kappa_{-}e^{i\psi/2} = 0
\end{equation}
Equations (\ref{eq:4_4_2c}) and (\ref{eq:4_4_3a}) involve only $\psi$ and $\hat{\rho}^{2}$ but not $R$. Up to factors of $\KK^2$, this system is the same as the system in \cite{D'Hoker:2016rdq}, and we will solve it with the same methods.  Adding twice (\ref{eq:4_4_2c}) to (\ref{eq:4_4_3a}) eliminates the term proportional to $e^{-i\psi}$, and we obtain,
\begin{equation}\label{eq:4_4_3d} 
\partial_{w}\ln \hat{\rho}^{2} + \frac{i}{2}\partial_{w}\psi - \partial_{w}\ln (1 - |\lambda|^{2}) +\hat{\rho}^{2}\kappa_{-}e^{i\psi/2} = 0
\end{equation}
Clearly, this equation involves only the following specific complex combination of $\hat{\rho}^{2}$ and $\psi$,
\begin{equation}
\label{eq:4_4_3e} 
K\xi = \hat{\rho}^{-2} e^{-i\psi/2}
\end{equation}
where we have included a factor of $K$ in the definition of $\xi$ for later convenience.
In terms of $\xi$ we may express (\ref{eq:4_4_3d}) as follows,
\begin{equation}
\label{eq:4_4_3f} 
K\partial_{w}\left( \xi (1 - |\lambda|^{2}) \right) = \kappa_{-} (1 - |\lambda|^{2}) = \kappa_{-} - \kappa_{+}\bar{\lambda} 
\end{equation}
where we have used the relation $\kappa_{+} = \lambda \kappa_{-}$.
The integrable structure of the system of equations (\ref{eq:4_4_1d}), (\ref{eq:4_4_1e}), and (\ref{eq:4_4_1f}) has therefore been exposed clearly with the help of this sequence of changes of variables. Indeed, equation (\ref{eq:4_4_3f}) involves only the field $\xi$, which is the combination of $\hat{\rho}$ and $\psi$ entering (\ref{eq:4_4_3e}). Having obtained $\xi$,  equation (\ref{eq:4_4_3d})  may be solved  for $\hat{\rho}$ and $\psi$. Finally, having $\hat{\rho}$ and $\psi$, equation (\ref{eq:4_4_2d}) becomes an equation for $R$ only, and we will see below that it can be solved as well.

\subsection{Solving for \texorpdfstring{$\psi$}{psi}, \texorpdfstring{$\hat{\rho}^{2}$}{rhohat2}, and \texorpdfstring{$R$}{R} in terms of \texorpdfstring{$\mathcal{A}_{\pm}$}{Apm}}

Having decoupled the reduced BPS equations in the preceding subsection, we will solve the decoupled equations in the present section. We begin by solving (\ref{eq:4_4_3f}) for $\xi$, then obtain $\psi$, $\hat{\rho}^{2}$, and $R$ as described above. We introduce locally holomorphic functions $\mathcal{A}_{\pm}$ such that,
\begin{equation}
\label{eq:4_5a} 
\kappa_{\pm} = K\partial_{w}\mathcal{A}_{\pm} 
\hskip 1in 
\lambda = \frac{\partial_{w}\mathcal{A}_{+}}{\partial_{w}\mathcal{A}_{-}}
\end{equation}
Given the one-forms $\kappa_{\pm}$, the functions $\mathcal{A}_{\pm}$ are unique up to an additive constant for each function.

\sm

With the conventions used to define  $\xi$ and $\cA_\pm$, the equations governing $\xi$ in terms of $\cA_\pm$ are identical to those of the $AdS_6 \times S^2$ case, and we import their solution  from \cite{D'Hoker:2016rdq}, 
\begin{equation}
\label{eq:4_5_1i} 
\xi = \frac{\mathcal{L}}{1 - |\lambda|^{2}}
\hskip 1in 
\mathcal{L} =  \mathcal{A}_{-} - \bar{\mathcal{A}}_{+} + \bar{\lambda}(\bar{\mathcal{A}}_{-} - \mathcal{A}_{+}) 
\end{equation}
Note that $\hat{\rho}$ and $\psi$ are directly determined by $\xi$ using equation (\ref{eq:4_4_3e}).

\sm

To solve for $R$, we begin with equation (\ref{eq:4_4_2d}) before using (\ref{eq:4_4_2c}) to eliminate the term proportional to $e^{-i\psi}$. We then divide the resulting equation by $R$, and find,
\begin{align}
\label{eq:4_5_2a} 
0 & = \left( \frac{1}{R^{2}} - 1 \right) \partial_{w}R + \left( R + \frac{1}{R} + 4\KK^2 \right) \left( i\partial_{w}\psi - \partial_{w}\ln (1 - |\lambda|^{2}) \right) 
\nonumber \\
& \qquad + \left( R + \frac{1}{R} + \KK^2 \right)(2\partial_{w}\ln\hat{\rho}^{2} - i\partial_{w}\psi)
\end{align}
Changing variable from $R$ to a new variable $W$, which we conveniently define by,
\begin{equation}
\label{eq:4_5_2b} 
\KK^2 W = R + \frac{1}{R}
\end{equation}
renders equation (\ref{eq:4_5_2a}) linear in $W$ with an inhomogeneous part,
\begin{equation}\label{eq:4_5_2c} 
\partial_{w}W - 2(W + 1)\partial_{w}\ln\hat{\rho}^{2} + (W + 1)\partial_{w}\ln (1 - \lambda\bar{\lambda}) = 3 i\partial_{w}\psi - 3 \partial_{w}\ln (1 - \lambda\bar{\lambda})
\end{equation}
We note that this equation is now independent of $\KK^2$ and therefore coincides with the corresponding equation for the $AdS_6 \times S^2$ case, whose solution we import from \cite{D'Hoker:2016rdq}, 
\begin{equation}
\label{eq:4_5_2d} 
W  = 2 + \frac{6 \kappa ^2 \, \cG}{|\p_w \cG|^2} 
\end{equation}
where we have defined $\kappa ^2 $ and $\cG$ through, 
\begin{align} 
\label{eq:4_5_2i} 
\kappa ^2 & = - |\p_w \cA_+|^2 + |\cA_-|^2 
\no \\
 \cG & = |\mathcal{A}_{+}|^2 - |\mathcal{A}_{-}|^2 + \mathcal{B} + \bar{\mathcal{B}}
  \no \\
 \partial_{w}\mathcal{B} & = 
 \mathcal{A}_{+}\partial_{w}\mathcal{A}_{-} - \mathcal{A}_{-}\partial_{w}\mathcal{A}_{+} 
\end{align}
Since $\p_w \cB$ is a holomorphic 1-form, there exists a locally holomorphic function $\mathcal{B}$, defined up to the addition of an arbitrary complex constant.  This completes the solution of the decoupled reduced BPS equations for the fields $\psi$, $\hat{\rho}$, and $R$.

\newpage

\section{Supergravity fields of the local solutions}\label{sec:sugra-sol-summary}

The general local half-BPS solution to Type IIB supergravity with $SO(2, 1) \oplus SO(7)$ symmetry can now be expressed in terms of the locally holomorphic functions $\cA_\pm$ introduced above. Here we will translate from the local solution of the reduced BPS equations to the supergravity fields and discuss some of the immediate properties of the solutions.

\sm

For comparison we present the $AdS_6\times S^2$ and $AdS_2 \times S^6$ cases  in parallel.
The five-form field strength and all fermion fields vanish. The remaining bosonic fields in both cases are distinguished merely by the parameter $\Lambda = \KK^2$. The metric Ansatz and $\Lambda$ are given by,
\begin{subequations}
\label{eq:5_2a}
\begin{align}
 ds^{2} & = f_{2}^{2} ds_{AdS_{2}}^{2} + f_{6}^{2} ds_{S^{6}}^{2} + ds_{\Sigma}^{2} &  \text{for $AdS_2\times S^6$} && \Lambda&=-1 
 \nonumber\\
 ds^{2} & = f_{6}^{2} ds_{AdS_{6}}^{2} + f_{2}^{2} ds_{S^{2}}^{2} + ds_{\Sigma}^{2} &  \text{for $AdS_6\times S^2$} & & \Lambda&=+1 
\end{align}
The remaining fields  in both cases are given by,
\begin{align} 
ds_{\Sigma}^{2} & = 4\rho^{2} dw d\bar{w} & F_{(3)} & = d \mathcal{C} \wedge \widehat{\rm vol}_2
\end{align}
\end{subequations}
where $\widehat{\rm vol}_2$ is the volume form on $AdS_2$ of unit radius for $\Lambda =-1$ and the volume form of $S^2$ with unit radius for the case $\Lambda =1$.  The metric functions $f_{2}$, $f_{6}$, $\rho$, and $\mathcal{C}$ and the dilaton-axion field $B$ are all functions on $\Sigma$.

\subsection{The metric functions}

The metric functions $f_{2}$, $f_{6}$, and $\rho$ are naturally expressed  in terms of composite quantities $\kappa ^2$ and $\cG$ defined in (\ref{eq:4_5_2i}), and the function $R$ obtained by eliminating $W$ between equations  (\ref{eq:4_5_2b}) and (\ref{eq:4_5_2d}). The latter is given in terms of $\kappa^2$ and $\cG$ by, 
\begin{equation}
\label{eq:5_2b} 
  \Lambda R + \frac{1}{\Lambda R} = 2 + 6\frac{\kappa^{2}\mathcal{G}}{|\partial_{w}\mathcal{G}|^{2}}
\end{equation}
To obtain the explicit expressions for the metric functions, we begin by eliminating $\alpha$ and $\beta$ from the combinations $(f_{6} \pm \tfrac{3}{\nu}f_{2})^2$ in favor of $\kappa_{\pm}$ and $f$ using (\ref{eq:3_8b}) and (\ref{eq:4_2f}), and we find, 
\begin{align}
\label{eq:E_1a} 
( f_{6} + \frac{3}{\nu}f_{2} )^{2} & = \frac{4c^{2}f^{2}}{\rho^{2}}|\kappa_{-}|^{2} |B\bar{\lambda} + 1|^{2} & ( f_{6} - \frac{3}{\nu}f_{2} )^{2} & = \frac{4c^{2}f^{2}}{\rho^{2}}|\kappa_{-}|^{2} |\lambda + B|^{2}
\end{align}
Changing variables from $B$ to $Z$ using (\ref{eq:4_4_1a}) and (\ref{eq:4_4_1b}), solving for $f_{2}$ and $f_{6}$, and expressing the result in terms of $|Z|^{2} = R$, we obtain,
\begin{align}
\label{eq:5_2c} 
f_{2}^{2} & = \frac{c^{2}\kappa^{2}(1 - \Lambda R)}{9\rho^{2}(1 + \Lambda R)} 
& 
f_{6}^{2} & = \frac{c^{2}\kappa^{2}(1 + \Lambda R)}{\rho^{2}(1 - \Lambda R)} 
\end{align}
To calculate $\rho^{2}$ we express the result of  (\ref{eq:4_4_2b}) for $\rho^2$ in terms of $\hat \rho^2$, use  (\ref{eq:4_4_3e}) to obtain $\hat \rho^2 $ in terms of $\xi$, and (\ref{eq:4_5_1i}) to express $\xi$ in terms of $\cL$, which in turn is given by, 
\begin{align}
\label{eq:E_1d} 
 \hat{\rho}^{4} & = \frac{1}{\xi\bar{\xi}} = \frac{(1 - \lambda\bar{\lambda})^{2}}{\mathcal{L}\bar{\mathcal{L}}}
 & \bar{\mathcal{L}}\partial_{w}\mathcal{A}_{-} & = -\partial_{w}\mathcal{G}
\end{align}
Expressing the result in terms of $R$, $\cG$ and $\kappa^2$, we find, 
\begin{align}
 \rho^2 = \frac{c (\kappa^2)^{{ 3 \over 2}} }{|\p_w \cG|} \frac{(R+\Lambda R^2)^\half }{(1-\Lambda R)^{ {3 \over 2}}} 
\end{align}
Alternatively, after making use of (\ref{eq:5_2b}), and eliminating $\rho^2$ from $f_2^2$ and $f_6^2$, we have,
\begin{align}
\label{eqn:metric}
 f_2^2&=\frac{c}{9}\sqrt{6\Lambda \cG}\left(\frac{1-\Lambda R}{1+\Lambda R}\right)^{\tfrac{3}{2}}
 &
 f_6^2&=c\sqrt{6\Lambda \cG} \left(\frac{1+\Lambda R}{1-\Lambda R}\right)^{\tfrac{1}{2}}
 &
 \rho^2&=\frac{c\kappa^2}{\sqrt{6\Lambda\cG}}\left(\frac{1+\Lambda R}{1-\Lambda R}\right)^{\tfrac{1}{2}}
\end{align}
Some care will be needed with the choice of the branch of the square root, which will be discussed in detail  in section \ref{sec:regularity}.

\subsection{The axion-dilaton}

The axion-dilaton field $B$ is obtained using its expression of (\ref{eq:4_4_1a}) in terms of  $Z$, eliminating $Z$ in favor of $R$ and $\psi$ using (\ref{eq:4_4_2a}), and eliminating $\psi$ in favor of $\xi$ and $\cL$ using (\ref{eq:4_5_1i}),  
\begin{equation}
\label{eq:E_1f} 
B = \frac{R e^{i\psi} - \lambda}{1 - \bar{\lambda}R e^{i\psi}} = \frac{K^{2} R \bar{\mathcal{L}} - \lambda\mathcal{L}}{\mathcal{L} - \bar{\lambda}K^{2} R \bar{\mathcal{L}}}
\end{equation}
Multiplying numerator and denominator by $|\kappa_{-}|^{2}$ and using (\ref{eq:E_1d}) to eliminate $\mathcal L$ yields, 
\begin{equation}
\label{eq:5_2d} 
B = \frac{\partial_w\cA_+\partial_{\bar{w}}\mathcal{G} - \Lambda R \, \partial_{\bar w}\bar\cA_- \partial_{w}\mathcal{G}}{\Lambda R \, \partial_{\bar w}\bar\cA_+ \partial_{w}\mathcal{G} - \partial_w\cA_-\partial_{\bar{w}}\mathcal{G}}
\end{equation}
One verifies that this field automatically satisfies $|B|<1$ provided $\kappa ^2 (1-R^2) >0$.

\subsection{Two-form and six-form flux potentials}
\label{sec:C2C6}

The evaluation of the two-form flux potential $C_{(2)}$ and of its magnetic dual six-form flux potential $C_{(6)}$ on the solutions to the BPS equations is considerably more involved than for the other supergravity fields. Here we shall summarize the result, and relay an account of the detailed calculations to appendix \ref{sec:flux}. As a byproduct, the calculations of the flux potentials will prove that the solutions to the BPS equations for 16 supersymmetries together with the Bianchi identities for the $P,Q$ one-forms, imply the Bianchi identity and field equation for the three-form field $G$.

\sm

Consider the three-form $F_{(3)}$ and dual seven-form $F_{(7)}$ field strengths defined by,
\begin{align}
F_{(3)} & =   f(G + B\bar{G})
\no \\
F_{(7)} & =  \star f (G - B \bar G) + { 4i \over 3} \left ( 2 C_{(4)} \wedge F_{(3)} -  F_{(5)} \wedge C_{(2)} \right )
\end{align}
where $\star$ denotes the Poincar\'e dual. It is a standard result that the  Bianchi identity for the field $F_{(3)}$ is given by $dF_{(3)}=0$. By inverting the relation between $F_{(3)}$ and $G$ one deduces the well-known Bianchi identity for $G $, which takes the form,
\bea
\label{BianG}
d G  - i Q \wedge G  + P \wedge \bar G  =0
\eea 
where  $P$ and $Q$ are given in terms of  $B$ by (\ref{PQG}). The field equation for $G$ is equivalent to the condition $d F_{(7)}=0$.  Here we shall be interested only in solutions to the field equations for which $F_{(5)}=C_{(2)} \wedge \bar F_{(3)}=0$, so that the field equation for $G$ reduces to,
\bea
\nabla ^P \left ( f ( G_{MNP} - B \bar G_{MNP} ) \right ) =0
\eea
The closure conditions on the three-form $F_{(3)}$ and on the seven-form $F_{(7)}$ may be solved locally in terms of flux potentials $C_{(2)}$ and $C_{(6)}$ by,
\begin{align}
dC_{(2)} & =   f(G + B\bar{G})
\no \\
dC_{(6)} & =  \star f (G - B \bar G)
\end{align}
In view of the $SO(2,1) \oplus SO(7)$ isometry algebra of $AdS_2 \times S^6$ and the $SO(5,2) \oplus SO(3)$ isometry algebra  of $AdS_6 \times S^2$, we have the following Ansatz for $C_{(2)}, C_{(6)}$ and $G$,
\begin{align}
C_{(2)} & = \cC \, \widehat{\rm vol}_2 & G & = g_{a}e^{a} \wedge {\rm vol}_2
\no \\
C_{(6)} & = \cM \, \widehat{\rm vol}_6
\end{align}
where $\widehat{\rm vol}_2$ and $\widehat{\rm vol}_6$ denote the volume forms of the maximally symmetric spaces of unit radius respectively of the two-dimensional and  six-dimensional factors of the space-time.
Integrating these equations for our solutions results in the following flux potentials,
\bea
\label{eq:5_2e} 
\mathcal{C} & = & \frac{4ic}{9}\Lambda\left\{ \frac{\partial_{\bar{w}}\bar{\mathcal{A}}_{-}\partial_{w}\mathcal{G} }{\kappa^{2}} - 2\Lambda R\frac{(\partial_{\bar{w}}\mathcal{G}\partial_{w}\mathcal{A}_{+} + \partial_{w}\mathcal{G}\partial_{\bar{w}}\bar{\mathcal{A}}_{-})}{(1 + \Lambda R)^{2}\kappa^{2}} - \bar{\mathcal{A}}_{-} - 2\mathcal{A}_{+}  \right\}
\no \\
 \cM & = &  
 -24 c^3\cG\left(\frac{\partial_w\cG\partial_{\bar w}\bar\cA_-}{\kappa^2}+3\bar\cA_-+2\cA_+\right)
 +80 c^3(\cW_++\bar\cW_-)
\no \\ && 
 +40 c^3 (\cA_++\bar\cA_-)(|\cA_+|^2-|\cA_-|^2)
\eea
where $\cW_\pm$ are locally holomorphic functions defined up to a constant by $\cA_\pm \p_w \cB = \p_w \cW_\pm$.

\subsection{\texorpdfstring{$SU(1,1)$}{SU(1,1)} transformations induced on the supergravity fields}\label{sec:su11}

The action of the global $SU(1,1)$ symmetry of Type IIB supergravity on the supergravity fields, as given in (\ref{eq:2_2a}),  is induced by an action of $SU(1,1)\otimes\C$ on $\cA_\pm$, in parallel with the $AdS_6 \times S^2$ case of \cite{D'Hoker:2016rdq}, 
\bea
\label{eq:5_3b} 
\mathcal{A}_{+} & \to & \mathcal{A}'_{+}  =  +u\mathcal{A}_{+} - v\mathcal{A}_{-} + a 
\no \\
\mathcal{A}_{-} & \to & \mathcal{A}'_{-}  =  -\bar{v}\mathcal{A}_{+} + \bar{u}\mathcal{A}_{-} + \bar a
\eea
where we have parametrized $SU(1,1)$ by $u, v \in \C$ with $|u|^{2} - |v|^{2} = 1$ and $a$ is a complex constant. The transformation of $\cB$ is given by, 
\begin{align}
\label{eq:5_3d} 
\mathcal{B} & \to \mathcal{B}'  = \mathcal{B} +  a\mathcal{A}'_{-} - \bar a\mathcal{A}'_{+}
\end{align}
These transformations leave $\kappa^2$ and $\cG$ and consequently also the metric functions invariant.
The condition $F_{(5)}=0$ is also left invariant.  
They transform $B$ as given in (\ref{eq:2_2a}), while $\cC$ and $\cM$ transform as follows,
\bea
\mathcal{C} & \to & \, \mathcal{C}'  ~ = u\mathcal{C} + v\bar{\mathcal{C}} - \mathcal C_0
\no \\
\cM & \to & \cM'  = u \cM - v \bar \cM + \cM_0
\eea
Consistently with the $SU(1,1)$ action of  (\ref{eq:2_2a}) and (\ref{eq:2_2b}), 
$F_{(3)}$ and $F_{(7)}$ transform  as follows,
\bea
F_{(3)} & \to & u F_{(3)} + v \bar F_{(3)}
\no \\
F_{(7)} & \to & u F_{(7)} - v \bar F_{(7)} 
\eea
where the first transformation law follows from the second equation in (\ref{eq:2_2a}) and  $F_{(3)}=dC_{(2)}$.

\newpage

\section{Verifying the field equations} \label{app:EOM}

Whether the BPS equations for 16 residual supersymmetries imply the full set of Bianchi identities and field equations for the form fields $P,Q,G,F_{(5)}$, the space-time metric, and the spin connection is, in general, an open problem. In our solution of the BPS equations, we have assumed the expression for the spin connection in terms of the metric, and we have assumed that the Bianchi identities for $P,Q$, given by, 
\begin{align}
0 & = dP - 2iQ\wedge P \nonumber \\
0 & = dQ + iP\wedge\bar{P} 
\end{align}
have been solved in terms of the axion-dilaton field $B$ by the first two equations in (\ref{PQG}). The Bianchi identity for the $F_{(5)}$ field is trivially satisfied since in our solutions we have $F_{(5)}=0$ as well as $G \wedge \bar G=0$. But the Bianchi identity for the field $G$ in (\ref{BianG}) was not assumed from the outset  and instead has been shown in subsection \ref{sec:C2C6} to result from the solution to the BPS equations and the Bianchi identity for $P,Q$. 

\sm

In this section we show that the field equations of Type IIB supergravity are obeyed for the general local solution obtained in section~\ref{sec:sugra-sol-summary}.  We continue to treat the $AdS_2\times S^6\times\Sigma$ and $AdS_6\times S^2\times\Sigma$ cases in parallel, and establish the Type IIB  field equations for both cases. In particular, we verify the field equations for the warped $AdS_6$ solutions obtained in \cite{D'Hoker:2016rdq}, a result that was not completely obtained in that paper.
The full Type IIB supergravity field equations for the bosonic fields are \cite{Schwarz:1983qr, Howe:1983sra},
\begin{align}
\label{eq:2_1c} 
0 & = \nabla^{M}P_{M} - 2iQ^{M}P_{M} + \frac{1}{24}G_{MNP}G^{MNP} \nonumber \\
0 & = \nabla^{P}G_{MNP} -iQ^{P}G_{MNP} - P^{P}\bar{G}_{MNP} + \frac{2}{3}i F_{(5)MNPQR}G^{PQR} \nonumber \\
0 & = R_{MN} - P_{M}\bar{P}_{N} - \bar{P}_{M}P_{N} - \frac{1}{6}(F_{(5)}^2)_{MN} \nonumber \\
& \qquad \qquad -\frac{1}{8}( G_{M}{}^{PQ} \bar{G}_{NPQ} + \bar{G}_{M}{}^{PQ}G_{NPQ} ) + \frac{1}{48}g_{MN}G^{PQR}\bar{G}_{PQR}
\end{align}
To show that these equations are satisfied, we start with Einstein's equations and then turn to the field equations for $B$ and the three-form flux. We will need the components of the Ricci tensor, which are derived for general $AdS_p\times S^q\times\Sigma$ warped products in appendix \ref{sec:Ricci}. We will use the labels $p$ and $q$ for the dimensions of the $AdS$ and $S$ parts of the geometry throughout this section, as well as $f_A$ and $f_S$ for their respective radii. The general procedure for verifying the field equations is to reduce them to a form where they only involve quantities for which we have given explicit expressions in terms of the holomorphic data in sec.~\ref{sec:sugra-sol-summary}, and then verify them via a strategy that will be explained in sec.~\ref{sec:eliminate-RG}.

\subsection{Einstein's equations}

For easier reference we reproduce here the explicit expressions for the components of the Ricci tensor along $\Sigma$, with $f_A$, $f_S$, $p$ and $q$ defined in appendix \ref{sec:Ricci},
\begin{align}
 R_{ww}&=
 -\frac{p}{f_A}\left[\partial_w-(\partial_w\ln\rho^2)\right]\partial_w f_A
 -\frac{q}{f_S}\left[\partial_w-(\partial_w\ln\rho^2)\right]\partial_w f_S
\nonumber\\
R_{w\bar w}&=-p\frac{\partial_w\partial_{\bar w}f_A}{f_A}-q\frac{\partial_w\partial_{\bar w}f_S}{f_S}
 -\partial_w\partial_{\bar w}\ln \rho^2 
\end{align}
We will also use the explicit expansions of $P$ and $G$,
\begin{align}\label{eq:PGexp}
 P&=p_z \rho dw+p_{\bar z}\rho d\bar w
 &
 G&=(g_zdw+g_{\bar z}d\bar w)\rho f_2^2\wedge \widehat{\rm vol}_2
\end{align}
where $\widehat{\rm vol}_2$ is the canonical volume form on $AdS_2$ of unit radius for $AdS_2\times S^6\times \Sigma$ and on $S^2$ of unit radius for $AdS_6\times S^2\times \Sigma$.

\subsubsection{Components along \texorpdfstring{$\Sigma$}{Sigma}}

The $ww$ component of Einstein's equations in (\ref{eq:2_1c}) simplifies to,
\begin{align}
 0&=R_{ww}-2P_w \bar P_w-\frac{1}{4}G_w^{\hphantom{w}PQ}\bar G_{wPQ}
\end{align}
Evaluating $G_w^{\hphantom{w}PQ}\bar G_{wPQ}$ amounts to contracting two volume forms on the two-dimensional space, for which the difference in signature between $AdS_2$ and $S^2$ is crucial. 
We thus find,
\begin{align}
\label{eq:G-contraction}
 G_w^{\hphantom{w}PQ}\bar G_{wPQ}&=2\Lambda \rho^2 g_z (g_{\bar z})^\ast
\end{align}
Using the expression for the components of $P$ in (\ref{eq:3_8a}) and that $\bar K/K=\Lambda$, the $ww$ components of Einstein's equations then become,
\begin{align}
 0&=R_{ww}-\frac{3\Lambda}{8}\rho^2g_z (g_{\bar z})^\ast
\end{align}
With (\ref{eq:E_2d}) this can be further evaluated to $ \rho^2g_z (g_{\bar z})^\ast=-16\Lambda f^4(\partial_w B)(\partial_w \bar B)$,
and the $ww$ component of Einstein's equations consequently becomes,
\begin{align}
\label{eqn:Einstein-ww}
 0&=R_{ww}+6 f^4(\partial_w B)(\partial_w \bar B)
\end{align}
Note that $f^4 d B  d \bar B$ is the Poincar\'e metric on the disc, and $\rm SU(1,1)$ invariant.

\sm

Using the expansions of $P$ and $G$ in (\ref{eq:PGexp}) and 
again that contractions of $G$ produce overall factors as given in (\ref{eq:G-contraction}), the $w\bar w$  component of Einstein's equations becomes,
\begin{align}
 0&=R_{w\bar w}-\rho^2\left[p_z(p_z)^\ast+p_{\bar z}(p_{\bar z})^\ast\right]-\frac{\Lambda}{4}\rho^2\left[g_z(g_z)^\ast+g_{\bar z}(g_{\bar z})^\ast\right]
 +\frac{1}{48}g_{w\bar w}G^{PQR}\bar G_{PQR}
\end{align}
We can now use $g_{w\bar w}=2\rho^2$ to evaluate,
\begin{align}\label{eq:GGbar}
 G^{PQR}\bar G_{PQR}&
 =3\Lambda\left[g_z(g_z)^\ast+g_{\bar z}(g_{\bar z})^\ast\right]
\end{align}
Using also (\ref{eq:3_8a}), the $w\bar w$ component of Einstein's equations consequently becomes,
\begin{align}
0&=R_{w\bar w}-\frac{\rho^2}{16}\left[\left(2\Lambda+\frac{\beta \bar\beta}{\alpha \bar\alpha}\right)|g_z|^2+\left(2\Lambda+\frac{\alpha \bar\alpha}{\beta \bar\beta}\right)|g_{\bar z}|^2\right]
\end{align}
Using (\ref{eq:E_2_1a}) with (\ref{eq:4_4_2a}) yields,
\begin{align}\label{eq:R-alphabeta}
 \frac{\alpha\bar \alpha}{\beta \bar \beta}&=R
\end{align}
With (\ref{eq:E_2d}), this yields for the $w\bar w$ component of Einstein's equations,
\begin{align}\label{eqn:Einstein-wwbar}
0&=R_{w\bar w}
-\left(2\Lambda R+1\right) f^4|\partial_w B|^2
-\left(\frac{2\Lambda}{R}+1\right)f^4|\partial_{ w} \bar B|^2
\end{align}

\subsubsection{Components along \texorpdfstring{$AdS_2\times S^6$ and $AdS_6\times S^2$}{AdS2xS6 and AdS6xS2}}

For the components of Einstein's equations in (\ref{eq:2_1c}) along the six-dimensional space, $AdS_6$ for the case of $AdS_6\times S^2$ and $S^6$ for the case of $AdS_2\times S^6$,
the only non-vanishing contributions are coming from the Ricci tensor and the last term. 
We thus find,
\begin{align}
 0&=R_{MN}+\frac{1}{48}g_{MN} G^{PQR}\bar G_{PQR}
 &
 \text{$M,N$ along $AdS_6$/$S^6$}
\end{align}
With (\ref{eq:GGbar}), (\ref{eq:E_2d}) and (\ref{eq:R-alphabeta}), we find,
\begin{align}
 \frac{1}{48} G^{PQR}\bar G_{PQR}
 &=\Lambda \rho^{-2}f^4\left[ R |\partial_w B|^2 + R^{-1}|\partial_w \bar B|^2\right]
\end{align}
For the components along the six-dimensional spaces, $AdS_6$ and $S^6$, we thus have,
\begin{align}\label{eqn:Einstein-AdS6-S6}
 0&=R_{mn}+\eta_{mn}\rho^{-2}f^4\left[ R |\partial_w B|^2 + R^{-1}|\partial_w \bar B|^2\right]
 &&\text{for $AdS_6$}
 \no\\
 0&=R_{ij}-\delta_{ij}\rho^{-2}f^4\left[ R |\partial_w B|^2 + R^{-1}|\partial_w \bar B|^2\right]
 &&\text{for $S^6$}
\end{align}
The three-form field $G$ has non-vanishing components along the two-dimensional space $AdS_2$/$S^2$,
and the corresponding components of Einstein's equations therefore become,
\begin{align}
 0&=R_{MN}-\frac{1}{8}\left(G_{M}^{\hphantom{M}PQ}\bar G_{NPQ}+\bar G_{M}^{\hphantom{M}PQ} G_{NPQ}\right)+\frac{1}{48}g_{MN} G^{PQR}\bar G_{PQR}
\end{align}
The contraction of $G$ in the last term has already been evaluated in (\ref{eq:GGbar}).
Explicitly evaluating the second term with  $M,N$ along $AdS_2$/$S^2$ produces a contribution similar to the last term, only with a different numerical coefficient.
A factor $\Lambda$ again originates from the difference in signature between $AdS_2$ and $S^2$.
The components of Einstein's equations along $AdS_2$/$S^2$ become,
\begin{align}
 0&=R_{MN}-\frac{3\Lambda}{16}g_{MN}\left( |g_z|^2+|g_{\bar z}|^2\right)
\end{align}
For the components along the two-dimensional spaces, $AdS_2$ and $S^2$, we thus have,
\begin{align}\label{eqn:Einstein-AdS2-S2}
 0&=R_{mn}+ 3\eta_{mn}\rho^{-2}f^4\left( R|\partial_w B|^2+R^{-1}|\partial_w \bar B|^2\right)
 & &\text{for $AdS_2$}
\no\\
 0&=R_{ij}-3\delta_{ij}\rho^{-2}f^4\left( R|\partial_w B|^2+R^{-1}|\partial_w \bar B|^2\right)
 & &\text{for $S^2$}
\end{align}

\subsection{Axion-dilaton field equations}

We now turn to the axion-dilaton equation. We will perform the index contractions as contractions of spacetime indices instead of frame indices, without introducing new notation.
The equation then reads,
\begin{align}
 0 & = \partial^M P_M - g^{MN}\Gamma_{MN}^R P_R - 2iQ^{M}P_{M} + \frac{1}{24}G_{MNP}G^{MNP}
\end{align}
With the definitions of $P$ and $Q$ in (\ref{PQG}), we find,
\begin{align}
 \partial^M P_M- 2iQ^{M}P_{M}
 &=
 2g^{w\bar w}\left( f^2 \partial_w \partial_{\bar w} B + 2 f^4 \bar B (\partial_w B)\partial_{\bar w} B\right)
\end{align}
The connection term in the covariant derivative evaluates to,
\begin{align}
 g^{MN}\Gamma_{MN}^R P_R
 &=
 -\frac{1}{2}g^{w\bar w}\left[P_w \partial_{\bar w}\ln(f_A^{2p}f_S^{2q})+P_{\bar w} \partial_{w}\ln(f_A^{2p}f_S^{2q})\right]
 \label{eq:gGammaP}
\end{align}
This leaves only the term involving $G$ to be evaluated. We find,
\begin{align}
 G_{MNQ}G^{MNQ}&=6\Lambda g_z g_{\bar z}
 =96\Lambda \rho^{-2}\frac{\alpha \bar\beta}{\bar \alpha \beta}f^4(\partial_w B)(\partial_{\bar w} B)
\end{align}
where (\ref{eq:E_2d}) was used to obtain the second equality.
Using (\ref{eq:R-alphabeta}) and (\ref{eq:4_2f}) shows,
\begin{align}
 \frac{\alpha \bar\beta}{\bar \alpha \beta}&=R \frac{\bar\beta^2}{\bar \alpha^2}=R \frac{\bar B \kappa_+ + \kappa_-}{\kappa_+ + B \kappa_-}
\end{align}
The complete equation of motion, after dividing by $2 g^{w\bar w}f^2$, becomes,
\begin{align}\label{eq:P-eom}
 0&=
 \partial_w \partial_{\bar w} B + 2 f^2 \bar B (\partial_w  B)\partial_{\bar w}B 
 \nonumber\\ &\hphantom{=}
 +\frac{p}{4 f_A^2}\left[ (\partial_{\bar w} B) \partial_w f_A^2 + (\partial_w B)\partial_{\bar w}f_A^2  \right]
 +\frac{q}{4 f_S^2}\left[ (\partial_{\bar w} B)\partial_w f_S^2 + (\partial_w B)\partial_{\bar w}f_S^2  \right]
 \nonumber\\ &\hphantom{=}
 +{4 \Lambda R} \frac{\bar B \kappa_+ + \kappa_-}{\kappa_+ + B \kappa_-}f^2(\partial_w B)(\partial_{\bar w} B)
\end{align}

\subsection{The 3-form flux field equation}

The field equation for the 3-form field $G$ with vanishing $F_{(5)}$ reads,
\begin{align}
\label{eq:G-eom}
 0 & = \nabla^{P}G_{MNP} -iQ^{P}G_{MNP} - P^{P}\bar{G}_{MNP}
\end{align}
We have already presented one proof that this field equations holds for our solution in section 5.3, by showing that the form $F_{(7)}$ is closed. Here, we provide a second proof, obtained by direct evaluation.

\sm

Analyzing (\ref{eq:G-eom}), we see that the last two terms vanish unless $M,N$ are both along $AdS_2$ for the $AdS_2\times S^6$ case or correspondingly along $S^2$ for the $AdS_6\times S^2$ case.  The only non-trivial components of the entire equation are when $M,N$ are either both on $S^2$/$AdS_2$, or one of them on $S^2$/$AdS_2$ and one on $\Sigma$. In the latter case, (\ref{eq:G-eom}) reduces to an equation that is satisfied automatically due to metric compatibility of the connection on $S^2$/$AdS_2$.  It therefore only remains to consider the case with both components on $S^2$/$AdS_2$. For notational convenience we will introduce coordinate indices $\mu$, $\nu$, which correspond to $AdS_2$ for the $AdS_2\times S^6$ case and to $S^2$ for the $AdS_6\times S^2$ case. We will also again perform index contractions as contractions of spacetime indices, without introducing additional notation.

\sm

When $M,N = \mu, \nu$ are both along $S^2$/$AdS_2$, the field equation  reads,
\begin{align}
 0&=\partial^P G_{\mu\nu P} -iQ^{P}G_{\mu\nu P} - P^{P}\bar{G}_{\mu\nu P}
 \nonumber\\&\hphantom{=}
 -g^{PQ}\Gamma_{PQ}^R G_{\mu\nu R}
 -g^{PQ}\left(\Gamma_{P\mu}^RG_{R\nu Q}+\Gamma_{P\nu}^RG_{\mu RQ}\right)
\end{align}
Evaluating the connection terms in analogy with (\ref{eq:gGammaP}), the field equation becomes, 
\begin{align}
 0&=
 \left(
 \partial_w -i Q_w  +\frac{1}{2}\partial_w \ln\frac{f_A^{2p}f_S^{2q}}{f_2^8}\right)G_{\bar w \mu\nu}
 -P_w \bar G_{\bar w \mu\nu}
 +(w\leftrightarrow \bar w)
\end{align}
We now use the expansion (\ref{eq:PGexp}) along with (\ref{eq:E_2d})  to replace,
\begin{align}
 G_{w\mu\nu}&= 4iK f_2^2 \frac{\alpha}{\beta}P_w \,\widehat{\rm vol}_{2\,\mu\nu}
 &
 G_{\bar w\mu\nu}&= -4iK\Lambda f_2^2 \frac{\bar \beta}{\bar \alpha}P_{\bar w}\,\widehat{\rm vol}_{2\,\mu\nu}
\end{align}
With  (\ref{eq:R-alphabeta}), we then find,
\begin{align}
  \frac{1}{4iK}\frac{\bar\alpha}{\bar \beta}\left(\partial_wG_{\bar w \mu\nu}+\partial_{\bar w}G_{w \mu\nu}\right) &=
 R\left(\partial_{\bar w}
 +\frac{1}{2}\partial_{\bar w}\ln\frac{\alpha^2}{\beta^2}\right)f_2^2 P_w \,\widehat{\rm vol}_{2\,\mu\nu}
 \nonumber\\ &\hphantom{=}
 -\Lambda\left(\partial_w 
 +\frac{1}{2}\partial_w\ln\frac{\bar\beta^2}{\bar\alpha^2}\right)f_2^2 P_{\bar w}\,\widehat{\rm vol}_{2\,\mu\nu}
 \nonumber\\
 \frac{1}{4iK}\frac{\bar\alpha}{\bar \beta}\left( P_w \bar G_{\bar w \mu\nu}+P_{\bar w}G_{w \mu\nu}\right)
 &=f_2^2\frac{\bar \alpha^2}{\bar\beta^2}\left( R^{-1}|P_{\bar w}|^2-\Lambda|P_w|^2\right)\,\widehat{\rm vol}_{2\,\mu\nu}
\end{align}
The equation of motion, after dividing by $4i\Lambda K f_2^2 \bar \beta/\bar \alpha$ and separating off the volume form on the two-dimensional space, consequently becomes,
\begin{align}
 0&=
 \Lambda R\left[\partial_{\bar w}-iQ_{\bar w}+\frac{1}{2}\partial_{\bar w}\ln\left(\frac{f_A^{2p}f_S^{2q}}{f_2^{4}}\frac{\alpha^2}{\beta^2}\right)\right]P_{w}
 \nonumber\\&\phantom{=}
 -\left[\partial_{w}-iQ_w+\frac{1}{2}\partial_w\ln\left(\frac{f_A^{2p}f_S^{2q}}{f_2^{4}}\frac{\bar\beta^2}{\bar\alpha^2}\right)\right]P_{\bar w}
 -\frac{\bar \alpha^2}{\bar\beta^2}\left(\frac{|P_{\bar w}|^2}{\Lambda R}-|P_w|^2\right)
\end{align}
Evaluating the derivatives  and using the components of $Q$ as defined in (\ref{PQG}) yields,
\begin{align}\label{eq:Geq}
 0&=(\Lambda R-1)\left(f^2\partial_w\partial_{\bar w}B+\frac{3}{2}\bar B P_w P_{\bar w}\right)
 +\left(\frac{1}{2}B\Lambda R+ \frac{\bar \alpha^2}{\bar\beta^2}\right)\left(|P_w|^2-\frac{|P_{\bar w}|^2}{\Lambda R }\right)
 \nonumber\\&\hphantom{=}+
 \frac{1}{2}\Lambda R P_{w}\partial_{\bar w}\ln\left(\frac{f_A^{2p}f_S^{2q}}{f_2^{4}}\frac{\alpha^2}{\beta^2}\right)
 -\frac{1}{2}P_{\bar w}\partial_w\ln\left(\frac{f_A^{2p}f_S^{2q}}{f_2^{4}}\frac{\bar\beta^2}{\bar\alpha^2}\right)
\end{align}
With $P_w=f^2\partial_w B$, $P_{\bar w}=f^2\partial_{\bar w} B$ as well as,
\begin{align}\label{eq:alphabar2dbetabar2}
 \frac{\bar\alpha^2}{\bar\beta^2}&=\frac{\partial_w\cA_+ + B \partial_w\cA_-}{\bar B \partial_w \cA_+ + \partial_w\cA_-}
\end{align}

\subsection{Explicitly evaluating the equations}\label{sec:eliminate-RG}

To summarize, the non-trivial components of Einstein's equations take the form given in (\ref{eqn:Einstein-ww}) for the $ww$ component,
in (\ref{eqn:Einstein-wwbar}) for the $w\bar w$ component, in (\ref{eqn:Einstein-AdS6-S6}) for the $AdS_6$/$S^6$ components,
and in (\ref{eqn:Einstein-AdS2-S2}) for the $AdS_2$/$S^2$ components.
The equation for the axion-dilaton scalar takes the form given in (\ref{eq:P-eom})
and the non-trivial components of the equation for $G$  are given in (\ref{eq:Geq}) with (\ref{eq:alphabar2dbetabar2}).

\sm

We will now describe the strategy to verify that these equations are satisfied. We use the explicit expressions for the metric functions in (\ref{eqn:metric}), for $B$ in (\ref{eq:5_2d}), and  for the components of the Ricci tensor. This reduces the field equations to a set of equations involving only the holomorphic functions and their derivatives, as well as $R$, $\cG$ and $\kappa^2$ along with their derivatives. We will avoid using the explicit definition for $\cG$ or $R$, since $\cG$ involves an integration that we have not performed for generic $\cA_\pm$  while the definition of $R$ involves a square root with a corresponding choice of branch that we do not wish to specify explicitly. The first step will be to make the expressions algebraic in $R$ and $\cG$, i.e.\ to eliminate all their derivatives. From the definitions for $\cG$ in (\ref{eq:4_5_2i}) and for $R$ in (\ref{eq:5_2b}), we straightforwardly derive,
\begin{align}
\partial_w R &=\frac{6\Lambda R^2}{R^2-1}\partial_w\left(\frac{\kappa^2\cG}{|\partial_w\cG|^2}\right)
\nonumber\\
\partial_w\cG&=\left(\bar \cA_+ - \cA_-\right)\partial_w\cA_+
+ \left(\cA_+ - \bar \cA_-\right)\partial_w\cA_-
\end{align}
The $\partial_{\bar w}$ derivatives of $R$ and $\cG$ are obtained by complex conjugation.
Repeatedly using these relations to reduce the rank in derivatives acting on $R$ and $\cG$, we can eliminate all derivatives of $\cG$ and $R$.
We now use the definition of $R$ in (\ref{eq:5_2b}) to eliminate $\cG$, by setting,
\begin{align}
 \cG&=\left(R+\frac{1}{R}-2\Lambda\right)\frac{|\partial_w\cG|^2}{6\Lambda \kappa^2}
\end{align}
Using also the explicit definition of $\kappa^2$, we have at this point reduced Einstein's equations to relations involving only the holomorphic functions and their differentials along with $R$. $\cG$ and its derivatives as well as the derivatives of $R$ are eliminated completely. Straightforward evaluation now shows that the equations are indeed satisfied for both, the $AdS_6$ and $AdS_2$ cases, with the corrsponding choices of $\Lambda$ and $K$ as well as of $p$ and $q$ for the dimensions of the $AdS$ and $S$ parts of the geometry.
This shows that the local solution to the BPS equations presented in sec.~\ref{sec:sugra-sol-summary} solves the field equations of Type IIB supergravity as well. We point out that, for the discussion of the BPS equations, $c$ was assumed real and $R$ constrained to be positive by its definition as absolute value of $Z$ in (\ref{eq:4_4_2a}), but that neither of these constraints appear to be necessary for the equations of motion to be satisfied.

\sm

We close this section with a comment on the sign of $G$. It is generally true that, for a solution to Type IIB supergravity, flipping the sign of $G$ produces another solution, since $G$ appears quadratically in the equations of motion. In general, supersymmetry is not preserved under this sign flip, since the BPS equations do depend on the sign of $G$. For our solutions, however, flipping the sign of $G$ again produces a supersymmetric solution. This may be seen from the fact that a sign reversal in $G$ corresponds to a special case of the $SU(1,1)$ transformations discussed in sec.~\ref{sec:su11}. Choosing $u=-1$ and $v=0$ indeed leaves all supergravity fields invariant except for the two-form potential $C_{(2)}$, on which it induces a sign flip that results in a sign reversal on $G$. The sign-flipped solution is therefore again in our class of supersymmetric solutions, although with a different form of the Killing spinors, which depend on the $\cA_\pm$ directly.


\newpage


\section{Reality, positivity, and regularity conditions}\label{sec:regularity}

The solutions obtained for the supergravity fields of the case  $AdS_2 \times S^6$ in the previous section satisfy the BPS equations, but are physically viable solutions only after certain reality, positivity, and regularity conditions are enforced on the supergravity fields of the  solutions. In this section we establish these conditions and uncover their implications on the functions $\cA_\pm$. We set $\Lambda =-1$ throughout this section.

\subsection{Reality and positivity conditions}

For an acceptable solution with appropriate signature, the metric is real-valued and the functions $f_2^2, f_6^2$ $\rho^2$ positive on $\Sigma$. There are no reality constraints on the fields $B$ and $C_{(2)}$, but $B$ is restricted by the condition of positive coupling constant, $|B|<1$.  We now extract the necessary and sufficient conditions on $\kappa ^2$, $\cG$, and $R$ for these properties to hold. 

\sm

Recalling that the functions $\kappa ^2$ and $\cG$, which were defined in (\ref{eq:4_5_2i}), are real-valued by construction,  and that $R$ is real and non-negative by construction, (\ref{eq:5_2b}) implies,  
\bea
\label{kG}
3\kappa ^2  \cG <-2|\partial_w\cG|^2
\eea
In particular, $\kappa^2$ and $\cG$ need to have opposite signs. Assuming positive $\rho^2$, the positivity of $f_2^2$ and $f_6^2$ in (\ref{eq:5_2c}) furthermore requires,
\bea
\label{kR}
  (1-R ) \, \kappa ^2  > 0
\eea
With this assumption $\rho^2$, given in (\ref{eq:5_2c}), is real and can be made positive by appropriately choosing the sign of the constant $c$.
To verify $|B| \leq 1$ we calculate $f^{2}$ using (\ref{eq:4_4_1b}),
\begin{equation}
\label{eq:5_4c} 
f^{2} = \frac{1}{1 - |B|^{2}} = 1 + \frac{|\lambda - Z^{2}|^{2}}{(1 - |\lambda|^{2})(1 - |Z|^{4})}
\end{equation}
Since $\kappa^{2}(1 - R) = |\kappa_{-}|^{2}(1 - |\lambda|^{2})(1 - |Z|^{2})$, (\ref{kR}) implies  $f^{2} \geq 1$. 
The reality and positivity conditions are therefore given by (\ref{kG}) and (\ref{kR}).

\subsubsection{Inversion and complex conjugation}

The space of allowed triplets $(\kappa^2, \cG, R)$  naturally divides into two branches, according to whether the conditions (\ref{kG}) and (\ref{kR}) are realized for $R>1$ or $R<1$. We shall refer to these branches as $\mB_\pm$, defined by,  
\bea
\label{eq:5_4h}
\mB_+ & = & \left \{ \kappa^{2} > 0, \qquad \cG < 0,   \qquad R < 1 \right \}
\no \\
\mB_- & = & \left \{ \kappa^{2} < 0, \qquad \cG > 0,   \qquad R > 1 \right \} 
\eea
These two branches are mapped into one another by an involution,
which is a combination of complex conjugation, reversal of the complex structure on $\Sigma$, and reversal of the indices~$\pm$ on the functions $\cA_\pm$, given by,
\begin{align}
\label{eq:5_4e}
\mathcal{A}_{\pm}(w) &\to \mathcal{A}'_{\pm}(w) = \bar{\mathcal{A}}_{\mp}(w) = \overline{\mathcal{A}_{\mp}(\bar{w})}
\end{align}
combined with $R\rightarrow R^{-1}$. 
This transformation leaves eq.~(\ref{eq:5_2b}) invariant and reverses the sign of $\kappa^2$ and $\cG$. 
It leaves the metric functions $f_2^2$, $f_6^2$ and $\rho^2$ invariant  and complex conjugates the fields $B$ and $\cC$,
\begin{align}
\label{eq:5_4g}
B(w,\bar{w}) \to B'(w,\bar{w}) & = \bar{B}(w,\bar{w}) = \overline{B(\bar{w},w)} \nonumber \\
\mathcal{C}(w,\bar{w}) \to \mathcal{C}'(w,\bar{w}) & = \bar{\mathcal{C}}(w,\bar{w}) = \overline{\mathcal{C}(\bar{w},w)}
\end{align}

\subsection{Global regularity and boundary conditions}

By inspection of the metric functions $f_2^2, f_6^2$, and $\rho^2$ in (\ref{eq:5_2c}), it is manifest that a supergravity solution considered on a compact subset $U$ of $\Sigma$ for which $(\kappa ^2, \cG, R)$ maps to a compact subset of either  $\mB_+$ or $\mB_-$ (but not both) is {\sl locally regular} in $U$. If the supergravity solution considered throughout a compact surface $\Sigma$ is such that $(\kappa ^2, \cG, R)$ maps to a compact subset of either branches $\mB_+$ or $\mB_-$ then the supergravity solution is {\sl globally regular} on $\Sigma$. 

\sm

If $\Sigma$ has a non-empty boundary, $\partial \Sigma$, additional regularity conditions have to be satisfied on $\partial\Sigma$. We assume geodesic completeness of the space-time manifold allowed for a supergravity solution, so that the boundary of space-time is at infinite geodesic distance (modulo issues of the Minkowski signature of the $AdS_2$-factor).  The only way we know how to realize this when $\Sigma$ has a boundary is by closing off the sphere $S^6$, namely $f_6^2 \to 0$, while keeping $f_2^2$ finite. In view of the expression obtained from (\ref{eq:5_2c}) for the ratio,
\bea
{f_6^2 \over f_2^2} = 9 { (1-R)^2 \over (1+R)^2}
\eea
this corresponds to the boundary condition $R=1$. As $R\rightarrow 1$, factors of $1-R$ in the expressions for the metric functions vanish at a number of places, and having a regular limit therefore imposes additional constraints: From the expression for $f^{2}_{2}$ in (\ref{eqn:metric}), we see that finiteness of the $AdS_2$ radius needs $\cG= \cO\left ( (1-R)^3 \right )$ as the boundary is approached. Similarly, from the finiteness of $\rho^{2}$ we then conclude that $\kappa ^2 = \cO(1-R)$. The boundary $\p \Sigma$ is therefore mapped to the common boundary of the two branches $\mB_\pm$, namely $\kappa ^2 = \cG =0$ and $R=1$. Lastly, in view of (\ref{eq:5_2b}) we also have the constraint that $\p_w \cG = \cO((1-R)^2)$.

\subsection{Implications of regularity and boundary conditions}
In this subsection we will discuss some immediate implications of the global regularity conditions for the structure of the solutions.

\subsubsection{No smooth solutions for compact \texorpdfstring{$\Sigma$}{Sigma} without boundary}

Assuming that $\cG$ is smooth, there are no globally regular solutions on a compact surface $\Sigma$ without boundary. The argument is parallel to the one already given for the case $AdS_6 \times S^2$. It is based on the following differential equation, 
\begin{equation}
\label{eq:5_5_4b}
\partial_{w}\partial_{\bar{w}}\mathcal{G} = -\kappa^{2}
\end{equation}
which readily follows from the definitions of $\kappa ^2 $ and $\cG$ in (\ref{eq:4_5_2i}). If $\cG$ is smooth, then on a compact surface without boundary, the integral of the left side over $\Sigma$ must vanish. But the sign of $\kappa^2$ is constant throughout $\Sigma$ so the integral of the right side cannot vanish, which is in contradiction to our assumptions. Hence such globally regular solutions cannot exist.
We are thus left with two options: either $\Sigma$ has a non-empty boundary, or $\Sigma$ is compact without boundary and the functions $\cA_\pm$ have singularities in $\Sigma$.

\subsubsection{No smooth solutions for compact \texorpdfstring{$\Sigma$}{Sigma} with boundary}

We show that for smooth $\cG$ and an arbitrary Riemann surface $\Sigma$ with non-empty boundary $\partial\Sigma$, the conditions $\mathcal{G}|_{\partial\Sigma} = 0$ and $\textrm{sgn}(\kappa^{2}) = -\textrm{sgn}(\mathcal{G})$ can not be satisfied simultaneously. We start from (\ref{eq:5_5_4b}) and  solve this equation along with the boundary condition $\mathcal{G}|_{\partial\Sigma} = 0$ to obtain the following integral equation,
\begin{equation}\label{eq:5_5_1b}
\mathcal{G}(w) = H(w) + \frac{1}{\pi}\int_{\Sigma}d^{2}z~G(w,z)\kappa^{2}(z)
\end{equation}
Here, $G(w,z)$ is the scalar Green function on $\Sigma$, which is symmetric $G(z,w) = G(w,z)$ and vanishes on the boundary $\partial\Sigma$,
\begin{align}\label{eq:5_5_1c}
\partial_{w}\partial_{\bar{w}}G(w,z) & = -\pi\delta(w,z) \nonumber \\
\left. G(w,z) \right|_{w \in \partial\Sigma} & = 0
\end{align}
As shown in detail in sec.~2.3 of \cite{DHoker:2017mds}, for any two points $w$, $z$ in the interior of $\Sigma$, the function $G(w,z)$ is strictly positive.
$H(w)$ is a harmonic function.
Since $\mathcal{G}(w)$ vanishes on the boundary by assumption, and by construction the Green function $G(w,z)$ vanishes for $w \in \partial\Sigma$, then $H$ itself must also vanish on $\partial\Sigma$, and we have
\begin{align}\label{eq:5_5_1d}
\partial_{w}\partial_{\bar{w}}H(w) & = 0 \nonumber \\
\left. H(w) \right|_{w \in \partial\Sigma} & = 0
\end{align}
By the min-max principle for harmonic functions, $H(w)$ takes its minimum and maximum values on the boundary of $\Sigma$.
Since $H(w)=0$ for $w \in \partial\Sigma$, this implies that $H(w) = 0$ both on $\partial\Sigma$ and in the interior of $\Sigma$. 
But this is incompatible with $\textrm{sgn}(\kappa^{2}) = -\textrm{sgn}(\mathcal{G})$, due to the positivity of the Green function $G(w,z)$,
which implies that the integral term in (\ref{eq:5_5_1b}) is strictly positive for the branch $\mB_+$ with $\kappa^2>0$ and strictly negative for the branch $\mB_-$ with $\kappa^2<0$.
Thus, no regular supergravity solutions exist when $\mathcal{G}$ vanishes on  $\p \Sigma$.


\newpage


\section{Double analytic continuation}\label{sec:doubleWick}

In this section we study the relation between $AdS_2\times S^6$ and $AdS_6\times S^2$ via double analytic continuation of the space-time manifold metrics in more detail, and discuss the implications from the perspective of the solutions to the BPS equations. At the level of the geometry, one may perform an analytic continuation from $AdS_6\times S^2$ to $AdS_2\times S^6$ via\footnote{The signs can be understood as follows. 
Starting from $AdS$ in mostly plus signature, one obtains Euclidean hyperbolic space by a standard Wick rotation in Poincar\'e coordinates. From Euclidean hyperbolic space, for which we can take global coordinates such that $ds^2=dr^2+\sinh^2\!r\, d\Omega^2$, we can then get to a sphere by setting $r=i\theta$. The resulting metric is $ds^2=-(d\theta^2+\sin^2\!\theta\, d\Omega^2)$, i.e.\ a sphere with negative signature.}
\begin{align}\label{eq:acont}
 AdS_6&\rightarrow - S^6 & S^2&\rightarrow - AdS_2
\end{align}
and these continuations can be extended straightforwardly to the remaining bosonic supergravity fields.
This does not produce a ten-dimensional spacetime of the desired signature, since the eight-dimensional symmetric space has changed signature from mostly plus to mostly minus while the metric on $\Sigma$ remains positive definite, but it does formally produce a solution to the equations of motion.
We may therefore wonder whether we can recover the analytic continuation of the globally regular $AdS_6\times S^2\times \Sigma$ solutions constructed in \cite{DHoker:2016ysh,DHoker:2017mds} as a special case of the $AdS_2\times S^6\times \Sigma$ solutions presented in sec.~\ref{sec:sugra-sol-summary}.
In the remainder of this section we will show that this is indeed the case, but that the result is neither regular (even leaving aside the signature issue) nor supersymmetric.

\sm

The construction in \cite{DHoker:2016ysh,DHoker:2017mds} started from $\Sigma$ a disc, realized as the upper half plane. The holomorphic functions $\cA_\pm$ were given by,
\begin{align}\label{eq:cA-AdS6}
 \cA_\pm &=\cA_\pm^0+\sum_{\ell=1}^L Z_\pm^\ell \ln(w-p_\ell)
 &
Z_+^\ell  &=
 \sigma\prod_{n=1}^{L-2}(p_\ell-s_n)\prod_{k \neq\ell}^L\frac{1}{p_\ell-p_k}
\end{align}
where $w$ is a complex coordinate on the upper half plane, $s_n$ a collection of points inside the upper half plane and $p_\ell$ a set of poles of the differentials $\partial_w\cA_\pm$ on the boundary of the upper half plane. 
With a suitable choice of the integration constant implicit in $\cG$, this produced,
\begin{align}\label{eq:wick2}
 \kappa^2,\cG &>0 \hskip 0.1in\text{on ${\rm int}(\Sigma)$}
 &
 \kappa^2&=\cG=0 \hskip 0.1in\text{on $\partial\Sigma$}
\end{align}
We can assume the same choice of holomorphic data as input for the $AdS_2\times S^6\times\Sigma$ solutions. The expressions for $\kappa^2$ and $\cG$ in terms of the locally holomorphic functions are the same for $AdS_2\times S^6\times\Sigma$ and $AdS_6\times S^2\times\Sigma$, such that we realize (\ref{eq:wick2}) in both cases.
Eq.~(\ref{eq:5_2b}) then implies $\Lambda R>0$ in the interior of the upper half plane and $\Lambda R\rightarrow 1$ on the boundary, and we choose the branch $0<\Lambda R\leq 1$.
This implies that $R$ is positive for $AdS_6\times S^2\times\Sigma$ and negative for $AdS_2\times S^6\times\Sigma$. 
We note that negative $R$ was not acceptable for solving the BPS equations, where $R$ was positive by construction. 
But, as noted at the end of sec.~\ref{app:EOM}, neither $R$ nor the constant $c$ are constrained by the equations of motion. So at the level of the equations of motion these $AdS_2\times S^6$ configurations are acceptable, and we have,
\begin{align}
 (\Lambda R)_{AdS_2\times S^6\times\Sigma}&=(\Lambda R)_{AdS_6\times S^2\times\Sigma}
\end{align}
The expressions for the supergravity fields in (\ref{eqn:metric}), (\ref{eq:5_2d}) and (\ref{eq:5_2e}) depend on $R$ only through this combination $\Lambda R$. The form of the axion-dilaton $B$ in (\ref{eq:5_2d}) is, in fact, exactly the same for $AdS_2\times S^6\times\Sigma$ and $AdS_6\times S^2\times\Sigma$. 
The metric functions in (\ref{eqn:metric}) are real provided that $c$ is chosen real for $AdS_6\times S^2\times\Sigma$ and imaginary for $AdS_2\times S^6\times\Sigma$, to compensate for the phase in $\sqrt{\Lambda\cG}$. They only differ between $AdS_2\times S^6\times\Sigma$ and $AdS_6\times S^2\times\Sigma$ through their signs:
while $f_2^2$, $f_6^2$ and $\rho^2$ all have the same sign for $AdS_6\times S^2\times\Sigma$, the sign of $f_2^2$ and $f_6^2$ is opposite to that of $\rho^2$ for $AdS_2\times S^6\times\Sigma$.
This is precisely as expected for solutions connected by the analytic continuation in (\ref{eq:acont}). The gauge potential $\cC$ in (\ref{eq:5_2e}) differs only by an overall factor of $i$ between $AdS_2\times S^6\times\Sigma$ and $AdS_6\times S^2\times\Sigma$. This produces the expected behavior under a Wick rotation for the three-form field strength, where one of the components along $S^2$ becomes timelike and picks up a factor of $i$. We have thus recovered the analytic continuation of the global $AdS_6\times S^2\times\Sigma$ solutions to $AdS_2\times S^6\times\Sigma$ via (\ref{eq:acont}), which is simply realized by the same choice of locally holomorphic functions.

\sm

This naive analytic continuation does, however, not lead to physically regular solutions. Aside from the inappropriate signs for the metric functions, it is still the two-dimensional space that collapses on the boundary of $\Sigma$. This was the desired behavior for the $AdS_6\times S^2\times\Sigma$ case, with the collapsing $S^2$ smoothly closing off spacetime. But it is not desirable for the $AdS_2\times S^6\times\Sigma$ solutions to have the $AdS_2$ cap off on $\partial\Sigma$.
Moreover, the solutions are not supersymmetric, since we do not recover them from the BPS equations where $R\geq 0$ was required by construction. The loss of supersymmetry under Wick rotation may be understood from the change in the Clifford algebra due to the changed signature in the two- and six-dimensional spaces.
The construction of physically regular supersymmetric solutions will therefore have to be revisited, with the regularity conditions and constraints spelled out in sec.~\ref{sec:regularity}.

\newpage

\section{Conclusion}\label{sec:conclusion}

We have constructed the general local form of solutions to Type IIB supergravity that are invariant under $SO(2,1)\oplus SO(7)$ and sixteen supersymmetries.
The geometry takes the form $AdS_2\times S^6$ warped over a two-dimensional Riemann surface $\Sigma$, and the local form of the solutions is strikingly similar to the $AdS_6\times S^2$ case considered in \cite{D'Hoker:2016rdq}. The entire solution is summarized, in parallel with the $AdS_6\times S^2$ case, in sec.~\ref{sec:sugra-sol-summary}, and we have verified for both cases that the solution to the BPS equations also satisfies the field equations of Type IIB supergravity in sec.~\ref{app:EOM}. The differences between the local solutions for $AdS_2\times S^6$ and $AdS_6\times S^2$ are subtle, and encoded entirely in sign flips at various places. 
\sm

To obtain physically acceptable solutions additional positivity and regularity conditions have to be imposed on the general local form of the solutions. We have presented a preliminary analysis of these conditions for $AdS_2\times S^6$ in sec.~\ref{sec:regularity}. Subtle but crucial differences between the solutions for $AdS_2\times S^6$ and $AdS_6\times S^2$ appear to render ineffective the strategy followed in \cite{DHoker:2017mds} to obtain global solutions for $AdS_6\times S^2$. An analytic continuation of the physically regular $AdS_6\times S^2$ solutions to $AdS_2\times S^6$, discussed in sec.~\ref{sec:doubleWick}, gives rise to field configurations which solve the field equations, but are neither regular nor supersymmetric. The construction of physically regular $AdS_2\times S^6$ solutions therefore poses an interesting new problem, which we hope to come back to in a follow-up paper.

\sm

One may speculate about the possible brane interpretation of warped $AdS_2\times S^6$ solutions. The physically regular $AdS_6 \times S^2$ solutions constructed in \cite{DHoker:2017mds} have a  compelling interpretation in terms of the conformal limit of $(p,q)$ 5-brane webs. A similar interpretation may be expected for physically regular $AdS_2$ solutions to Type IIB supergravity  in terms of webs of $(p,q)$ strings. It will be interesting to see whether $AdS_2$ solutions with this interpretation  can be found within our $SO(2,1)\oplus SO(7)$ invariant Ansatz.

\sm

Finally, we comment on the superalgebra structure of our $AdS_2 $ solutions. While the five-dimensional superconformal algebra is unique and corresponds to a specific real form of $F(4)$, there exist several superconformal algebras for $AdS_2$ \cite{VanProeyen:1986me,DHoker:2008wvd}. They are $SU(1,1|4), OSp(8|2,\R)$, and $OSp(4^*|4)$, with maximal bosonic subalgebras respectively  realized by $AdS_2 \times S^5 \times S^1 \times \Sigma$, $AdS_2 \times S^7 \times L$, and $AdS_2 \times S^2 \times S^4 \times \Sigma$, where $\Sigma$ is a Riemann surface and $L$ is a one-dimensional line. The last case was solved already in \cite{D'Hoker:2007fq}.

\section*{Acknowledgements}
We are happy to thank Michael Gutperle, Juan Maldacena and Andrea Trivella for useful discussions and comments. The work of all three authors is supported in part by the National Science Foundation under grant PHY-16-19926. ED  also gratefully acknowledges support by a Fellowship from the Simons Foundation.


\newpage

\appendix

\section{Clifford algebra basis adapted to the Ansatz}\label{sec:Clifford}

The signature of the space-time metric is chosen to be $(- + \cdots +)$. The Dirac-Clifford algebra is defined by $\{\Gamma^{A},\Gamma^{B} \} = 2\eta^{AB}I_{32}$, where $A, B$ are the 10-dimensional frame indices. We construct a basis for the Clifford algebra that is well-adapted to $AdS_{2} \times S^{6} \times \Sigma$ Ansatz, with the frame labeled as in (\ref{eq:2_3d}),
\begin{align}
\Gamma^{m} &= \gamma^{m} \otimes I_{8} \otimes I_{2} && m = 0,1 \nonumber \\
\Gamma^{i} &= \gamma_{(1)} \otimes \gamma^{i} \otimes I_{2} && i =  2, 3, 4, 5, 6, 7 \nonumber \\
\Gamma^{a} &= \gamma_{(1)} \otimes \gamma_{(2)} \otimes \gamma^{a} && a = 8, 9
\end{align}
where a convenient basis for the lower dimensional Dirac-Clifford algebra is as follows,
\begin{align}
\gamma^{0} & = i\sigma^{2} & \gamma^{2} &= \sigma^{1} \otimes I_{2} \otimes I_{2} && \nonumber \\
\gamma^{1} & = \sigma^{1} , & \gamma^{3} &= \sigma^{2} \otimes I_{2} \otimes I_{2} && \nonumber \\
&& \gamma^{4} &= \sigma^{3} \otimes \sigma^{1} \otimes I_{2} && \nonumber \\
&& \gamma^{5} &= \sigma^{3} \otimes \sigma^{2} \otimes I_{2} && \nonumber \\
&& \gamma^{6} &= \sigma^{3} \otimes \sigma^{3} \otimes \sigma^{1} & \gamma^{8} & = \sigma^{1} \nonumber \\
&& \gamma^{7} &= \sigma^{3} \otimes \sigma^{3} \otimes \sigma^{2} , & \gamma^{9} & = \sigma^{2}
\end{align}
We will also need the chirality matrices on the various components of $AdS_{2} \times S^{6} \times \Sigma$,
\begin{align}
\Gamma^{01} & = \gamma_{(1)} \otimes I_{8} \otimes I_{2}  & \gamma_{(1)}  & = \sigma^{3}
\nonumber \\
\Gamma^{234567} & = -iI_{2} \otimes \gamma_{(2)} \otimes I_{2}  & \gamma_{(2)} & = \sigma^{3} \otimes \sigma^{3} \otimes \sigma^{3}
\nonumber \\
\Gamma^{89} & = iI_{2} \otimes I_{8} \otimes \gamma_{(3)} & \gamma_{(3)} & = \sigma^{3}
\end{align}
The 10-dimensional chirality matrix in this basis is then given by,
\begin{equation}
\Gamma^{11} = \Gamma^{0123456789} = \gamma_{(1)} \otimes \gamma_{(2)} \otimes \gamma_{(3)} = \sigma^{3} \otimes \sigma^{3} \otimes \sigma^{3} \otimes \sigma^{3} \otimes \sigma^{3}
\end{equation}
The complex conjugation matrices in each component are defined as follows,
\begin{align}
(\gamma^{m})^{*} &= + B_{(1)}\gamma^{m}B^{-1}_{(1)} && \left( B_{(1)} \right)^{*} B_{(1)} = +I_{2} && B_{(1)} = I_{2} \nonumber \\
(\gamma^{i})^{*} &= - B_{(2)}\gamma^{i}B^{-1}_{(2)} && \left( B_{(2)} \right)^{*} B_{(2)} = +I_{8} && B_{(2)} = \sigma^{2} \otimes \sigma^{1} \otimes \sigma^{2} \nonumber \\
(\gamma^{a})^{*} &= - B_{(3)}\gamma^{a}B^{-1}_{(3)} && \left( B_{(3)} \right)^{*} B_{(3)} = -I_{2} && B_{(3)} = \sigma^{2}
\end{align}
where in the last column we have also listed the form of these matrices in our particular basis. The 10-dimensional complex conjugation matrix $\mathcal{B}$ satisfies,
\begin{align}
(\Gamma^{M})^{*} &=  \mathcal{B}\Gamma^{M}\mathcal{B}^{-1} & \mathcal{B}^{*} \mathcal{B} & = I_{32} & [\mathcal{B},\Gamma^{11}] & = 0
\end{align}
and in this basis is given by,
\begin{align}
\mathcal{B}  = -iB_{(1)} \otimes \left( B_{(2)}\gamma_{(2)} \right) \otimes B_{(3)}
 = I_{2} \otimes \sigma^{1} \otimes \sigma^{2} \otimes \sigma^{1} \otimes \sigma^{2}
\end{align}

\section{Derivation of the BPS equations}\label{sec:BPS}

In reducing the BPS equations, we will use the following decompositions of $\varepsilon$ and $\mathcal{B}^{-1}\varepsilon^{*}$,
\begin{align}
\varepsilon & = \sum_{\eta_{1},\eta_{2}}\chi^{\eta_{1},\eta_{2}}\otimes\zeta_{\eta_{1},\eta_{2}} & \mathcal{B}^{-1}\varepsilon^{*} = & \sum_{\eta_{1},\eta_{2}}\chi^{\eta_{1},\eta_{2}}\otimes\star\zeta_{\eta_{1},\eta_{2}}
\end{align}
where we use the abbreviations,
\begin{align}\label{eq:stardef} 
\star\zeta_{\eta_{1},\eta_{2}} & = -i\sigma^{2}\eta_{2}\zeta^{*}_{\eta_{1},-\eta_{2}} & \star\zeta & = \tau^{(02)}\otimes\sigma^{2}\zeta^{*}
\end{align}
in $\tau$-matrix notation. We will also need the chirality relations,
\begin{align}
\sigma^{3}\zeta_{\eta_{1},\eta_{2}} & = -\zeta_{-\eta_{1},-\eta_{2}} & \tau^{(11)}\otimes\sigma^{3}\zeta & = -\zeta
\end{align}

\subsection{The dilatino equation}

Reduced to the Ansatz, and using the above decomposition, the dilatino equation is,
\begin{align}
0 & = iP_{A}\Gamma^{A}\mathcal{B}^{-1}\varepsilon^{*} - \frac{i}{24}\Gamma\cdot G\varepsilon \nonumber \\
& = ip_{a}\Gamma^{a}\sum_{\eta_{1},\eta_{2}}\chi^{\eta_{1},\eta_{2}}\otimes\star\zeta_{\eta_{1},\eta_{2}} - \frac{i}{4}g_{a}\Gamma^{a}\sum_{\eta_{1},\eta_{2}}\chi^{-\eta_{1},\eta_{2}}\otimes\zeta_{\eta_{1},\eta_{2}}
\end{align}
where we have used the following simplifications for the inner products,
\begin{align}
P_{A}\Gamma^{A} & = p_{a}\Gamma^{a} \nonumber \\
\Gamma \cdot G & = 3! \, g_{a}\Gamma^{01a} = 6g_{a}\Gamma^{a} \gamma_{(1)}\otimes I_{8}\otimes I_{2}
\end{align}
Using the expression for $\star\zeta$ and reversing the sign of $\eta_{2}$, we extract  an equation satisfied by the $\zeta$-spinors, and recover the reduced dilatino BPS equation announced in (\ref{eq:3_3a}).

\subsection{The gravitino equation}

The gravitino equation is,
\begin{align}
0 & = (d + \omega)\varepsilon - \frac{i}{2}Q\varepsilon + \mathfrak{g}\mathcal{B}^{-1}\varepsilon^{*} \nonumber \\
\omega & = \frac{1}{4}\omega_{AB}\Gamma^{AB} \nonumber \\
\mathfrak{g} & = -\frac{1}{96}e_{A}\left( \Gamma^{A}(\Gamma\cdot G) + 2(\Gamma\cdot G)\Gamma^{A}  \right)
\end{align}
where $A$, $B$ are the 10-dimensional frame indices.

\subsubsection{The calculation of \texorpdfstring{$(d + \omega)\varepsilon$}{(d+omega)eps}}

The spin connection components for a generic space-time of the form $AdS_{p}\times S^{q}\times\Sigma$ are worked out in appendix \ref{sec:Ricci}. Here we quote results for the case of $p=2$ and $q=6$, and in particular we reproduce equation (\ref{eq:spincon}),
\begin{align}
\omega^{m}{}_{n} & = \hat{\omega}^{m}{}_{n} & \omega^{m}{}_{a} & = e^{m}D_{a}\ln f_{2} \nonumber \\
\omega^{i}{}_{j} & = \hat{\omega}^{i}{}_{j} & \omega^{i}{}_{a} & = e^{i}D_{a}\ln f_{6}
\end{align}
which gives the relevant spin connection components, along with $\omega^{a}{}_{b}$ whose explicit form we will not need. The hats refer to the canonical connections on $AdS_{2}$ and $S^{6}$, respectively. We replace the covariant derivative of the spinor along the $AdS_{2}$ and $S^{6}$ directions by the corresponding group action, $SO(2,1)$ and $SO(7)$, as defined in (\ref{eq:3_1a}). The additional term that appears in going from $\nabla$ to $\hat{\nabla}$ is due to the warp factors in the ten-dimensional metric. The covariant derivatives along $AdS_{2}$ and $S^{6}$, respectively, are given by,
\begin{align}
& (m) & \nabla_{m}\varepsilon & = \left( \frac{1}{f_{2}}\hat{\nabla}_{m} + \frac{D_{a}f_{2}}{2f_{2}}\Gamma_{m}\Gamma^{a} \right)\varepsilon \nonumber \\
& (i) & \nabla_{i}\varepsilon & = \left( \frac{1}{f_{6}}\hat{\nabla}_{i} + \frac{D_{a}f_{6}}{2f_{6}}\Gamma_{i}\Gamma^{a} \right)\varepsilon
\end{align}
as well as $\nabla_{a}\varepsilon$ along $\Sigma$. Using the Killing spinor equations (\ref{eq:3_1a}) to eliminate the hatted covariant derivatives, as well as the equation $\Gamma^{a} = \gamma_{(1)}\otimes\gamma_{(2)}\otimes\gamma^{a}$, we have,
\begin{align}
& (m) & \nabla_{m}\varepsilon & = \Gamma_{m}\sum_{\eta_{1},\eta_{2}}\chi^{\eta_{1},\eta_{2}}\otimes \left( \frac{1}{2f_{2}}\eta_{1}\zeta_{\eta_{1},\eta_{2}} + \frac{D_{a}f_{2}}{2f_{2}}\gamma^{a}\zeta_{-\eta_{1},-\eta_{2}} \right) \nonumber \\
& (i) & \nabla_{i}\varepsilon & = \Gamma_{i}\sum_{\eta_{1},\eta_{2}}\chi^{\eta_{1},\eta_{2}}\otimes \left( \frac{i}{2f_{6}}\eta_{2}\zeta_{-\eta_{1},\eta_{2}} + \frac{D_{a}f_{6}}{2f_{6}}\gamma^{a}\zeta_{-\eta_{1},-\eta_{2}} \right)
\end{align}
As we will show, each term in the gravitino equation contains $\Gamma_{A}\chi^{\eta_{1},\eta_{2}}$, which we argue are linearly independent. Therefore, we will require that the coefficients vanish independently along the various directions of $AdS_{2}$, $S^{6}$, and $\Sigma$.

\subsubsection{The calculation of \texorpdfstring{$\mathfrak{g}\mathcal{B}^{-1}\varepsilon^{*}$}{gBepsilon}}

The relevant expression is as follows,
\begin{equation}
\mathfrak{g}\mathcal{B}^{-1}\varepsilon^{*} = -\frac{3!}{96}e_{B}g_{a} ( \Gamma^{B}\Gamma^{01a} + 2\Gamma^{01a}\Gamma^{B} )\mathcal{B}^{-1}\varepsilon^{*}
\end{equation}
We make use of the following identities,
\begin{align}
\Gamma^{m}\Gamma^{01b} + 2\Gamma^{01b}\Gamma^{m} & = 3\Gamma^{m}\Gamma^{01b} \nonumber \\
\Gamma^{i}\Gamma^{01b} + 2\Gamma^{01b}\Gamma^{i} & = -\Gamma^{i}\Gamma^{01b} \nonumber \\
\Gamma^{a}\Gamma^{01b} + 2\Gamma^{01b}\Gamma^{a} & = \Gamma^{01}(3\delta^{ab} - \Gamma^{ab}) = (\gamma_{1}\otimes I_{8}\otimes I_{2})(3\delta^{ab}I_{2} - \gamma^{ab})
\end{align}
where $\gamma^{ab} \equiv i\varepsilon^{ab}\sigma^{3}$ and $\varepsilon^{89} = +1$. Projecting along the various directions, we obtain,
\begin{align}
& (m) & & \Gamma_{m}\sum_{\eta_{1},\eta_{2}}\chi^{\eta_{1},\eta_{2}}\otimes \left( -\frac{3}{16}g_{a}\gamma^{a}\star\zeta_{\eta_{1},-\eta_{2}} \right) \nonumber \\
& (i) & & \Gamma_{i}\sum_{\eta_{1},\eta_{2}}\chi^{\eta_{1},\eta_{2}}\otimes \left( \frac{1}{16}g_{a}\gamma^{a}\star\zeta_{\eta_{1},-\eta_{2}} \right) \nonumber \\
& (a) & & \sum_{\eta_{1},\eta_{2}}\chi^{\eta_{1},\eta_{2}}\otimes \left( -\frac{3}{16}g_{a}\star\zeta_{-\eta_{1},\eta_{2}} + \frac{1}{16}g_{b}\gamma^{ab}\star\zeta_{-\eta_{1},\eta_{2}} \right)
\end{align}

\subsubsection{The complete gravitino BPS equation}

We now assemble the reduced gravitino equations. Requiring the coefficients of $\Gamma_{A}\chi^{\eta_{1},\eta_{2}}$ to vanish independently, then rewriting the relations using the $\tau$-matrix notation, we have,
\begin{align}
& (m) & 0 & = \frac{1}{2f_{2}}\tau^{(30)}\zeta + \frac{D_{a}f_{2}}{2f_{2}}\tau^{(11)}\gamma^{a}\zeta - \frac{3}{16}g_{a}\tau^{(01)}\gamma^{a}\star\zeta \nonumber \\
& (i) & 0 & = \frac{i}{2f_{6}}\tau^{(13)}\zeta + \frac{D_{a}f_{6}}{2f_{6}}\tau^{(11)}\gamma^{a}\zeta + \frac{1}{16}g_{a}\tau^{(01)}\gamma^{a}\star\zeta \nonumber \\
& (a) & 0 & = \left( D_{a} + \frac{i}{2}\hat{\omega}_{a}\sigma^{3} - \frac{i}{2}q_{a} \right)\zeta - \frac{3}{16}g_{a}\tau^{(10)}\star\zeta + \frac{1}{16}g_{b}\tau^{(10)}\gamma^{ab}\star\zeta
\end{align}
where $\hat{\omega}_{a} = (\hat{\omega}_{89})_{a}$ is the spin connection along $\Sigma$, and we have used the fact $\Gamma^{89} = i\sigma^{3}$. Eliminating the star using the definition (\ref{eq:stardef}), then multiplying the $(m)$ and $(i)$ equations by $\tau^{(11)}$, we recover the system of reduced gravitino BPS equations announced in (\ref{eq:3_3b}).

\section{Calculation of the flux potentials}\label{sec:flux}
\setcounter{equation}{0}

In this appendix, we present the  calculations of the reduced flux potentials $\cC$ and $\cM$.

\subsection{Calculation of the flux potential $\cC$}

Expressing the field strength $G$ in terms of $g_a$, the equations for the derivatives of the potential $\cC$ are given by,
\begin{align}
\label{eq:E_2c} 
\partial_{w}\mathcal{C} & = \rho f_{2}^{2}f(g_{z} + B\bar{g}_{z}) 
\no \\
\partial_{\bar{w}}\mathcal{C} & = \rho f_{2}^{2}f(g_{\bar{z}} + B\bar{g}_{\bar{z}})
\end{align}
Converting $G$ into $P$ and then into derivatives of $B$ using (\ref{PQG}) and (\ref{eq:3_8a}) yields,
\begin{align}\label{eq:E_2d} 
\rho g_{z} & = 4iK\frac{\alpha}{\beta}f^{2}\partial_{w}B & \rho g_{\bar{z}} & = -4i K^{3}\frac{\bar{\beta}}{\bar{\alpha}}f^{2}\partial_{\bar{w}}B
\end{align}
along with $\bar{g}_{z}=(g_{\bar{z}})^\ast$ and $\bar{g}_{\bar{z}}=(g_{z})^\ast$,
and we obtain the following expressions,
\begin{align}\label{eq:E_2e} 
\partial_{w}\mathcal{C} & = 4iKf_{2}^{2}f^{3}\left( \frac{\alpha}{\beta}\partial_{w}B + B\frac{\beta}{\alpha}\partial_{w}\bar{B} \right) \nonumber \\
\partial_{w}\bar{\mathcal{C}} & = 4iKf_{2}^{2}f^{3}\left( \frac{\beta}{\alpha}\partial_{w}\bar{B} + \bar{B}\frac{\alpha}{\beta}\partial_{w}B \right)
\end{align}
We will now apply the same changes of variables used to solve the BPS equations. In the derivation of $\cC$ as well as $\cM$, it will be useful to have the derivatives of $B$ and $\bar B$ expressed in terms of the new variables, and these derivatives are given by,
\begin{align}
\label{derB}
\p_w B & = { 1 - |\lambda|^2 \over (1- \bar \lambda Z^2)^2} \p_w Z^2 - { \p _w \lambda \over 1- \bar \lambda Z^2}
\no \\
\p_w \bar B & = { 1 - |\lambda|^2 \over (1- \lambda \bar Z^2)^2} \p_w \bar Z^2 + { \bar Z^2 (\bar Z^2 - \bar \lambda) \p _w \lambda \over (1- \lambda \bar Z^2)^2}
\end{align}

\subsubsection{Expressing variables in terms of holomorphic functions}

Recall that we have,
\begin{align}\label{eq:E_2_1a} 
\frac{\alpha}{\beta} & = \left( \frac{\bar{\lambda} + \bar{B}}{\bar{\lambda}B + 1} \right)^{\tfrac{1}{2}} = \bar{Z}\left( \frac{1 - \bar{\lambda}Z^{2}}{1 - \lambda\bar{Z}^{2}} \right)^{\tfrac{1}{2}} , & \left| \frac{\alpha}{\beta} \right| & = |Z|
\end{align}
as well as the expressions for $B$ in (\ref{eq:4_4_1a}) and $f^{2}$ in (\ref{eq:4_4_1b}). Using these, along with the rearrangement formula following from (\ref{eq:4_4_3e}),
\begin{equation}\label{eq:E_2_1d} 
\frac{K}{\hat{\rho}^{2}\bar{Z}|Z|} = \frac{\bar{\xi}}{|Z|^{2}}
\end{equation}
we write (\ref{eq:E_2e}) as,
\begin{align}
\label{eq:E_2_1e} 
\partial_{w}\mathcal{C} & = \frac{4ic}{9}\frac{(1 - K^{2}|Z|^{2})}{(1 + K^{2}|Z|^{2})^{3}}\frac{\bar{\xi}}{|Z|^{2}}\left( \partial_{w}|Z|^{4} - \lambda(\partial_{w}\bar{Z}^{2} + \bar{Z}^{4}\partial_{w}Z^{2}) - \bar{Z}^{2}(1 - |Z|^{4})\partial_{w}\lambda \right) 
\nonumber \\
\partial_{w}\bar{\mathcal{C}} & = \frac{4ic}{9}\frac{(1 - K^{2}|Z|^{2})}{(1 + K^{2}|Z|^{2})^{3}}\frac{\bar{\xi}}{|Z|^{2}}\left( -\bar{\lambda}\partial_{w}|Z|^{4} + \partial_{w}\bar{Z}^{2} + \bar{Z}^{4}\partial_{w}Z^{2} \right)
\end{align}
Next, we define the combination $\bar{\mathcal{C}} + \bar{\lambda}\mathcal{C}$ by its derivative with respect to $w$ as,
\begin{equation}\label{eq:E_2_1f} 
\partial_{w}(\bar{\mathcal{C}} + \bar{\lambda}\mathcal{C}) = \frac{4iK^{2}c}{9}\mathcal{P}_w
\end{equation}
where $\mathcal{P}_w$ is given by,
\begin{equation}
\label{eq:E_2_1g} 
\mathcal{P}_w = \xi\frac{(1 - K^{2}|Z|^{2})}{(1 + K^{2}|Z|^{2})^{3}} \left( (1 - |\lambda|^{2})(\partial_{w}\ln\bar{Z}^{2} + \bar{Z}^{2}\partial_{w}Z^{2}) + (1 - |Z|^{4})\partial_{w}(1 - |\lambda|^{2}) \right)
\end{equation}
Using $\xi(1 - |\lambda|^{2}) = \mathcal{L}$ to eliminate $\xi$ and then changing variables via $Z^{2} = R e^{i\psi}$, we find,
\begin{equation}
\label{eq:E_2_1i} 
\mathcal{P}_w = \mathcal{L}\frac{(1 - K^{2} R)}{(1 + K^{2} R)^{3}} \left\{ \left( R + \frac{1}{R} \right)\partial_{w}R - (1 - R^{2})\left( i\partial_{w}\psi - \partial_{w}\ln (1 - |\lambda|^{2}) \right) \right\}
\end{equation}
The combinations of $R$ and $\partial_{w}W$ can be expressed in terms of $W$ defined in (\ref{eq:4_5_2b}) and its derivative $\partial_{w}W$. The latter is given by equation (\ref{eq:4_5_2c}), from which we eliminate $\hat{\rho}$ in favor of $\xi$, as well as eliminate $\xi$ in favor of $\mathcal{L}$ and $\lambda$, to obtain,
\begin{equation}\label{eq:E_2_1k} 
\partial_{w}W = -(W + 1)\partial_{w}\ln \mathcal{L}\bar{\mathcal{L}} + (W - 2)\partial_{w}\ln (1 - |\lambda|^{2}) + 3 i\partial_{w}\psi
\end{equation}
Using the relation $e^{i\psi} = K^{2}\bar{\mathcal{L}}/\mathcal{L}$ to express the derivative $i\partial_{w}\psi$ in terms of $\mathcal{L}$, then separating the dependence of $\mathcal{L}$ and $\bar{\mathcal{L}}$, we find,
\begin{equation}
\label{eq:E_2_1n} 
\mathcal{P}_w = \frac{2\mathcal{L}}{(W + 2)^{2}} [ (W^{2} + 2 W - 2)\partial_{w}\ln\mathcal{L} + (W - 2)\{ \partial_{w}\ln(1 - |\lambda|^{2}) - \partial_{w}\ln\bar{\mathcal{L}} \} ]
\end{equation}
The last term inside the square brackets can be replaced using the differential equation for $W$ in (\ref{eq:E_2_1k}), which upon expressing $\psi$ in terms of $\mathcal{L}$ takes the form,
\begin{equation}
\label{eq:E_2_1o} 
(W - 2)\{ \partial_{w}\ln (1 - |\lambda|^{2}) - \partial_{w}\ln\bar{\mathcal{L}} \} = \partial_{w}W + (W + 4)\partial_{w}\ln\mathcal{L}
\end{equation}
Eliminating this term between (\ref{eq:E_2_1k}) and (\ref{eq:E_2_1n}) and integrating the result, we find,
\begin{equation}\label{eq:E_2_1q} 
\mathcal{P}_w = \partial_{w}\left( \frac{2\mathcal{L}(W + 1)}{W + 2} \right)
\end{equation}
Therefore, we have, with some holomorphic function $\mathcal{K}_{1}$,
\begin{equation}
\label{eq:E_2_1r} 
\bar{\mathcal{C}} + \bar{\lambda}\mathcal{C} = \frac{4iK^{2}c}{9}\left( \frac{2\mathcal{L}(W + 1)}{W + 2} + \bar{\mathcal{K}}_{1} \right)
\end{equation}
Proceeding analogously for $\bar{\mathcal{C}}$, and using the equations for $\partial_{w}\xi$ and $\partial_{w}\bar{\xi}$, we find,
\begin{equation}\label{eq:E_2_1t} 
\bar{\mathcal{C}} = \frac{4iK^{2}c}{9} \left( -2\frac{\xi + \bar{\lambda}\bar{\xi}}{W + 2} + \frac{\mathcal{L}}{1 - |\lambda|^{2}} + \mathcal{A}_{-} + \bar{\mathcal{K}}_{2} \right)
\end{equation}
for some holomorphic function $\mathcal{K}_{2}$. To determine $\mathcal{K}_{2}$, we equate the two different expressions for $\bar{\mathcal{C}} + \bar{\lambda}\mathcal{C}$, eliminate the dependence on $W$  and separate holomorphic and anti-holomorphic dependences to obtain,
\begin{equation}\label{eq:E_2_1w} 
\mathcal{C} = \frac{4iK^{2}c}{9} \left( +2\frac{\bar{\xi} + \lambda\xi}{W + 2} - \frac{\bar{\mathcal{L}}}{1 - |\lambda|^{2}} - (\bar{\mathcal{A}}_{-} + 2\mathcal{A}_{+} + \mathcal{K}_{0}) \right)
\end{equation}
where $\mathcal{K}_{0}$ is an arbitrary complex constant. 
Eliminating $\xi$ and $\cL$ leads to  (\ref{eq:5_2e}).

\subsection{Calculation of the reduced flux potential $\cM$}

To evaluate $\cM$, we proceed in analogy with the calculation of $\cC$.  Reducing $G_{(3)}$ to $P$ and then expressing this quantity in terms of derivatives of $B$ using (\ref{PQG}) and (\ref{eq:3_8a}) yields,
\begin{align}
\p_w \cM & = 4K f_6^6 f^3 \left ( - { \alpha \over \beta} \p_w B + B { \beta \over \alpha} \p_w \bar B \right )
\no \\
\p_w \bar \cM & = 4K f_6^6 f^3 \left ( - {  \beta \over \alpha } \p_w \bar B + \bar B { \alpha \over \beta } \p_w B \right )
\end{align}
Using (\ref{eq:E_2_1a}) and (\ref{derB}) to express $\alpha/\beta$ and the derivatives of $B$ and $\bar B$ in terms of $Z$, $\lambda$ and $\hat\rho^2$, as well as the following expression,
\bea
f_6^6 f^3 = { c ^3 \over \hat \rho^6} \, { |1- \bar \lambda Z^2|^3 \over |Z|^3}
\eea
we find, after some simplifications, 
\bea
\p_w \cM & = & 
4K { c^3 \bar Z  \over \hat \rho^6 |Z|^3}  \Big  ( - (1 - |\lambda|^2) (1 - \lambda \bar Z^2) \p_w Z^2
+ (1 - |\lambda|^2) (Z^2 - \lambda)   \p_w \ln \bar Z^2 
\no \\ & & \hskip 0.6in 
 + |1-\lambda \bar Z^2|^2 \p_w \lambda  + |Z^2-\lambda|^2 \p_w \lambda \Big  )
 \no \\
 \p_w \bar \cM & = & 
4K { c^3 \bar Z  \over \hat \rho^6 |Z|^3}  \Big  ( - (1 - |\lambda|^2) (1 - \bar \lambda Z^2) \p_w \ln \bar Z^2
+ (1 - |\lambda|^2) (\bar Z^2 - \bar \lambda)   \p_w Z^2 
\no \\ & & \hskip 0.6in 
 -2 (\bar Z^2 - \bar \lambda )(1-\bar \lambda Z^2)  \p_w \lambda \Big  )
 \label{eq:dM}
\eea
Significant cancellations occur when combining these results into the following form,
\bea
\p_w ( \bar \cM - \bar \lambda \cM) =
4K { c^3 \bar Z^3  \over \hat \rho^6 |Z|^3} (1 - |\lambda|^2)^3 \left (  { \p_w (Z^2+ \bar Z^{-2}) \over 1 - |\lambda|^2} 
+ { \bar \lambda (Z^2 + \bar Z^{-2}) -2 \over (1-|\lambda|^2)^2}\partial_w\lambda \right )
\eea
Combining the relations,
\bea\label{c29}
K \xi = { e^{-i \psi /2} \over \hat \rho^2} 
\hskip 0.7in 
{ \bar Z \over |Z|} = e^{- i \psi /2}
\hskip 0.7in 
\xi (1 - |\lambda|^2) = \cL
\eea
we find, 
\bea\label{c30}
{  \bar Z^3  \over \hat \rho^6 |Z|^3} (1 - |\lambda|^2)^3 = K^3\cL^3
\eea
Recognizing a total derivative in the expression for $\p_w ( \bar \cM - \bar \lambda \cM)$, we may rearrange the result as follows,
\bea
\p_w ( \bar \cM - \bar \lambda \cM) = 4c^3 \cL^3 \, 
\p_w \left ( { Z^2 + \bar Z^{-2} - 2 \bar \lambda ^{-1}  \over 1-|\lambda|^2 }  \right )
\eea
Extracting a total derivative again significantly simplifies the expression, and we have,
\bea\label{eq:MMbar}
\p_w ( \bar \cM - \bar \lambda \cM) = 12c^3\cS_w +
\p_w \left [ 4c^3 \cL^3  \left ( { Z^2 + \bar Z^{-2} - 2 \bar \lambda ^{-1} \over 1-|\lambda|^2 } \right ) \right ]
\eea
where $\cS_w$ is given by,
\bea
\cS_w = -  \left ( { Z^2 + \bar Z^{-2} - 2 \bar \lambda ^{-1} \over 1-|\lambda|^2 }  \right )\cL^2 \p_w \cL
\eea
Using the relation $\p_w \cL = (1 - |\lambda|^2) \p_w \cA_-$, we find a remarkable cancellation of the factor $(1-|\lambda|^2)$, and recover an expression which,
after converting $Z$ and $\bar Z$ to $R$ and $\psi$ using (\ref{eq:4_4_2a}), is given as follows,
\bea
\cS_w & = & - \cL^2 e^{i \psi} \left (R + { 1 \over R} \right ) \p_w \cA_- + { 2\cL^2 \over \bar \lambda} \p_w \cA_-
\eea
Next, we make use of the identity $\cL e^{i \psi} = \Lambda \bar \cL$, and then use $\p_w \cG = -\bar \cL \p_w \cA_-$ to obtain the following expression, 
\bea
\cS_w = - \left (\Lambda R + { 1 \over \Lambda R} \right ) { |\p_w \cG|^2 \over \pbw \bar \cA_-} + { 2\cL^2 \over \bar \lambda}\p_w \cA_-
\eea
Using the definition of $R$ as well as $\kappa^2=-\partial_w\partial_{\bar w}\cG$, this can be rewritten as follows,
\begin{align}\label{eq:Sw}
 \cS_w&=\partial_w\left(6 \frac{\cG\partial_{\bar w}\cG}{\partial_{\bar w}\bar\cA_-}\right)+8\cL \partial_w\cG +{ 2\cL^2 \over \bar \lambda}\p_w \cA_-
\end{align}
The first term already is in integrated form, and the last two terms are 
of degree at most two in $\cA_+$ and $\cA_-$ and of degree at most one in $\p_w\cA_+$ and $\p_w\cA_+$, and they may be integrated in their holomorphic dependence on $w$.

\subsubsection{Integrating for \texorpdfstring{$\bar\cM-\bar\lambda\cM$}{Mb-lambdab M}}

We introduce two locally holomorphic functions, $\cW_\pm$  such that
\begin{align}
\label{eq:cW}
 \partial_w\cW_\pm =\cA_\pm \partial_w\cB
\end{align}
One may then verify by straightforward evaluation that the second term in (\ref{eq:Sw}) can be integrated as follows
\begin{align}\label{eq:Sw-2}
 \cL\partial_w\cG&=-
 \partial_w\left[\left(\frac{1}{2}\cG+\cB\right)(\bar\cA_+-\bar\lambda\bar\cA_-)-\frac{1}{2}\left(|\cA_+|^2-|\cA_-|^2\right)\cL
 +\bar\lambda\cW_+-\cW_-\right]
\end{align}
For the remaining term we use repeatedly the relation $\partial_w\cL=\partial_w\cA_- - \bar\lambda\partial_w\cA_+$ and find
\begin{align}\label{eq:Sw-3}
 \cL^2 \p_w \cA_-
 = \partial_w\Big[&
 \cL^2\cA_-+(\bar\lambda \cA_+-\cA_-)\cA_-\left(\cL+\frac{\bar\lambda\cA_+-\cA_-}{3}\right)
 \nonumber\\
 &+\bar\lambda\left(\bar\cA_+-\bar\lambda\bar\cA_-\right)\cB+\frac{2}{3}\bar\lambda^2\cW_+-\frac{2}{3}\bar\lambda\cW_- \Big]
\end{align}
Combining (\ref{eq:Sw-2}) and (\ref{eq:Sw-3}) with the expression for $S_w$ in (\ref{eq:Sw}) shows that we have thus integrated $S_w$.
Coming back to the expression for $\partial_w(\bar \cM - \bar \lambda \cM)$ in (\ref{eq:MMbar}), we can thus integrate for $\bar \cM - \bar \lambda \cM$. After a number of simplifications and using that $\bar\lambda\bar\cL-\cL=(1-|\lambda|^2)(\bar\cA_+-\cA_-)$, we find
\begin{align}
\label{eq:cM0}
 \frac{\bar \cM - \bar \lambda \cM}{4c^3}&=
 \frac{2\cL^2}{\bar\lambda}(\bar\cA_+-\cA_-)
 -12\left( \cG- |\cA_+|^2+|\cA_-|^2\right)\cL
 +\frac{2\cA_-}{\bar\lambda}(\bar\lambda \cA_+-\cA_-)^2
 \nonumber\\&\hphantom{=}\,
 -6\left(\bar\cA_+-\bar\lambda\bar\cA_-\right)\left(3\cB+2\cG+\frac{\cL\cA_-}{\bar\lambda}\right)
  -20(\bar\lambda\cW_+-\cW_-)
  +\bar\cV_1
\end{align}
where $\cV_1$ is a so far arbitrary locally holomorphic function. The terms of degree $-1$ in $\bar\lambda$ combine to a purely anti-holomorphic function and can be absorbed into $\cV_1$.

\subsubsection{Integrating \texorpdfstring{$\bar\cM$}{cMb}}

To fix the so far unspecified locally holomorphic function, we go back to the expression for $\partial_w\bar\cM$ in (\ref{eq:dM}).
Using (\ref{c30}), and combining all the derivative terms, it can be rewritten as
\begin{align}
 \partial_w\bar\cM=-4c^3\cL^3\bar\lambda^{-1}\partial_w\left(\frac{(1-\bar\lambda Z^2)(1-\bar\lambda\bar Z^{-2})}{(1-|\lambda|^2)^2}\right)
\end{align}
To evaluate the terms in the derivative further, we perform the variable changes to $R$ and $\psi$ and then to $\cL$, which yields
\begin{align}
 (1-\bar\lambda Z^2)(1-\bar\lambda \bar Z^{-2})&=
 1-\frac{\bar\lambda\bar\cL}{\cL}\left(\Lambda R+\frac{1}{\Lambda R}\right)+\frac{\bar\lambda^2\bar\cL^2}{\cL^2}
 \nonumber\\
 &=
 \cL^{-2}\left(\cL-\bar\lambda\bar\cL\right)^2-6\bar\lambda\frac{\bar \cL}{\cL}\frac{\kappa^2\cG}{|\partial_w\cG|^2}
\end{align}
Using that $\bar\lambda\bar\cL-\cL=(1-|\lambda|^2)(\bar\cA_+-\cA_-)$ and $\partial_w\cG=-\bar\cL\partial_w\cA_-$, yields
\begin{align}
 \frac{(1-\bar\lambda Z^2)(1-\bar\lambda \bar Z^{-2})}{1-|\lambda|^2}&=\cL^{-2}(\bar\cA_+-\cA_-)^2-\frac{6\bar\lambda\cG\cL^{-2}}{1-|\lambda|^2} 
\end{align}
We thus have
 \begin{align}
 \partial_w\bar\cM=-4c^3\cL^3\partial_w\left(\bar\lambda^{-1}\cL^{-2}(\bar\cA_+-\cA_-)^2-\frac{6\cG\cL^{-2}}{1-|\lambda|^2} \right)
\end{align}
Extracting a total derivative and using again that $\partial_w\cL=(1-|\lambda|^2)\partial_w\cA_-$, 
this expression can be further evaluated. Extracting total derivatives iteratively eventually yields
\begin{align}
 \frac{\partial_w\bar\cM}{4c^3}&=\partial_w\left(\frac{6\cL\cG}{1-|\lambda|^2}+(\cA_+-\bar\cA_-)(\bar\cA_+-\cA_-)^2-18\cA_-\cG\right)
 \nonumber\\ &\hphantom{=}\,
 +18\cA_-\partial_w\cG-3(\bar\cA_+-\cA_-)^2\partial_w\cA_+
\end{align}
The terms in the second line are again of degree at most two in $\cA_\pm$ and at most one in $\partial_w\cA_-$, and can be integrated straightforwardly.
We find
\begin{align}
 18\cA_-\partial_w\cG-3(\bar\cA_+-\cA_-)^2\partial_w\cA_+
 =
 \partial_w\Big[& 12|\cA_+|^2\cA_- -12 \bar\cA_+\cB - 9|\cA_-|^2\cA_- 
 \nonumber\\&
 - 3|\cA_+|^2\bar\cA_+ + 20 \cW_- -\cA_+\cA_-^2\Big]
\end{align}
This completes the integration and we conclude that
\begin{align}\label{eq:cM1}
 \bar\cM=4c^3\Big[&
 \frac{6\cL\cG}{1-|\lambda|^2}+(\cA_+-\bar\cA_-)(\bar\cA_+-\cA_-)^2-18\cA_-\cG
 +12|\cA_+|^2\cA_-
 \nonumber\\&
   -12 \bar\cA_+\cB - 9|\cA_-|^2\cA_- 
 - 3|\cA_+|^2\bar\cA_+ + 20 \cW_- -\cA_+\cA_-^2 + \bar\cV_2
 \Big]
\end{align}
with a locally holomorphic function $\cV_2$.
To fix $\cV_1$ and $\cV_2$, we equate the two expressions for $\cM$ in (\ref{eq:cM1}) and (\ref{eq:cM0}) and isolate the holomorphic and anti-holomorphic dependences.
This yields
\begin{align}
 \cV_2&=20\cW_+-12\cA_+\cB+\cA_-\cA_+^2
\end{align}
The final form of $\cM$ then becomes
\begin{align}
 \cM=8c^3\Big[&
 3\cG\left(\frac{\partial_w\cG}{\bar\lambda \partial_w\cA_+-\partial_w\cA_-}-3\bar\cA_--2\cA_+\right)
 +10(\cW_++\bar\cW_-)
 \nonumber\\&
 +5(\cA_++\bar\cA_-)(|\cA_+|^2-|\cA_-|^2)
 \Big]
\end{align}
where $\cW_\pm$  are defined up to complex constants by (\ref{eq:cW}).

\section{Curvature for \texorpdfstring{$AdS_p\times S^q\times\Sigma$}{AdSpxSqxSigma} warped products}\label{sec:Ricci}
\setcounter{equation}{0}

To compute the curvature for the $AdS_6 \times S^2$ and $AdS_2 \times S^6$ cases in parallel,
we generalize the setting to $AdS_p \times S^q$ warped over a Riemann surface $\Sigma$, with metric
\bea
ds^2 = 4 \rho^2 |dw|^2 + f_A ^2 \, d\hat s^2 _{AdS_p}+ f_S ^2 \, d\hat s^2 _{S^q} 
\eea
The functions $\rho^2, f_A^2, f_S^2$ depend only on $\Sigma$ and $d\hat s^2 _{AdS_p}$, $d\hat s^2 _{S^q} $ are respectively the $SO(2,p-1)$ and  $SO(q+1)$-invariant metric of unit radius on $AdS_p$ and $S^q$.

\sm

With $\hat e^m$ and $\hat e^i$ denoting the orthonormal frames for the unit radius $AdS_p$ and $S^q$, respectively,
we make the following choice for the orthonormal frame,
\bea
e^m & = & f_A \, \hat e^m \hskip 1in m=1,\cdots, p
\no \\
e^i &=& f_S \, \hat e^i \hskip 1.1in i=p+1, \cdots, p+q
\eea
combined with  $e^z$, $e^{\bar z}$ for $\Sigma$ as defined in (\ref{eq:4_2b}).
We collectively denote the frame indices by $A,B$. 
The frame metrics are given as in (\ref{eq:3_4m}) for $\Sigma$ and $\eta _{mn} = {\rm diag} \, [ - + \cdots +]$, $\delta _{ij} = {\rm diag} \, [ + \cdots +]$.
Denoting the connection 1-form by $\om ^A{}_B$, the torsion equations are,
\bea
d e^A + \om ^A {}_B \wedge e^B & = & 0
\eea
The connection forms of the  symmetric spaces, denoted by a hat in analogy to the notation for the frame, are defined by the analogous vanishing torsion conditions.
The connection components on $\Sigma$ are given by
\bea
\label{eq:wzz}
\om^z {}_z  =  dw \, \p_w \ln \rho - d\bar w \, \pbw \ln \rho
\eea
We use the notation of (\ref{eq:4_2b}) for the frame-covariant derivative when no connection is needed,
and analogous notation when a connection is needed, e.g.\ $D_z v^a=\rho^{-1}(\partial_w v^a+\omega_{w\,\hphantom{a}b}^{\hphantom{w\,}a}v^b)$. 
The remaining components are then,
\bea
\label{eq:spincon} 
\omega ^m {}_n = \hat \omega ^m {}_n 
& \hskip 1in & \omega ^m {}_a = e^m D_a \ln f_A 
\no \\
\omega ^i  {}_j \, = \, \hat \omega ^i  {}_j \,
& \hskip 1in & \omega ^i {}_a \, = \, e^i D_a \ln f_S 
\eea
and $\omega ^m {}_i = 0$.
The components of the Riemann tensor, defined via
\begin{align}
 \frac{1}{2}R^A_{\hphantom{A}BCD} e^C\wedge e^D&=d\omega^A_{\hphantom{A}B}+\omega^A_{\hphantom{A}C}\wedge\omega^C_{\hphantom{C}B}
\end{align}
are then found as
\begin{align}
R^m{}_{nAB} & =  -\frac{1+|D_a f_A |^2}{f_A^2}
\,\delta ^m{}_{[A} \, \eta _{B]n}
&
R^i{}_{jAB} & =   {1-|D_a f_S |^2 \over f_S^2} 
\,\delta ^i{}_{[A} \, \delta _{B]j}
\no \\
R^a{}_{bAB} & =  R^{(2)} \,\delta ^a{}_{[A} \, \delta _{B]b}
&
R^m{}_{iAB} & =  -   {D_a f_A \, D^a f_S \over f_A f_S} \,\delta ^m{}_{[A} \, \delta _{B]i}
\no \\
R^m{}_{aAB} & =  -   {D_b D_a f_A  \over f_A } \,\delta ^b{}_{[A} \, \delta_{B]}{}^m
&
R^i{}_{aAB} & =  -   {D_b D_a f_S  \over f_S } \delta ^b{}_{[A} \, \delta _{B]}{}^i
\end{align}
where we use the notation $|D_a \ln f |^2 \equiv (D^a \ln f) (D_a \ln f)$
and square brackets denote antisymmetrization of the enclosed indices, e.g.\ 
$\delta ^m{}_{[A} \, \eta _{B]n}=\delta ^m{}_A \, \eta _{Bn} - \delta ^m{}_B \, \eta _{An}$.
Furthermore,
\bea
R^{(2)} = - { 1 \over \rho^2} \p_w \pbw \ln \rho
\eea
where the normalization is such that $R^{(2)}=-1$ when $\rho^2 = y^{-2}$ corresponds to the Poincar\'e metric on the upper half space.
The components of the Ricci tensor (in frame index convention) are defined by
$R_{AB} = R^C_{\hphantom{C}ACB}$, and the non-vanishing ones are given by,
\bea
R_{mn} & = & \eta _{mn} \left (
- {p-1 \over f_A^2} - (p-1) { |D_a f_A |^2 \over f_A ^2} - q {D^a f_A D_a f_S \over f_A f_S}
- {D^a D_a f_A \over f_A}  \right )
\no \\
R_{ij} & = & \delta _{ij} \left (
+ {q-1 \over f_S^2} -  (q-1) { |D_a f_S |^2 \over f_S ^2} - p {D^a f_A D_a f_S \over f_A  f_S}
- {D^a D_a f_S \over f_S}  \right )
\no \\
R_{ab} & = & 
- p { D_b D_a f_A \over f_A} 
- q { D_b D_a f_S \over f_S}  + R^{(2)} \delta _{ab}
\label{eq:Rab}
\eea

\bibliographystyle{JHEP.bst}
\bibliography{ads2}
\end{document}